\documentclass[twocolumn]{aastex63}

\usepackage{graphicx,epstopdf,amsmath}
\usepackage{import}
\usepackage{tipa}

\usepackage{color}

\newcommand{\phpl}{PhPl}

\newcommand{\jcoph}{JCoPh}

\newcommand{\jgra}{JGRA}
\begin{document}

\title{A Model for the Coupled Eruption of a Pseudostreamer and Helmet Streamer}
%\title{Coupled Pseudostreamer and Helmet Streamer Eruption}

%\title{Coupled Helmet Streamer--Pseudostreamer eruptions}

%\title{Coupled Eruptions from Large Pseudostreamers}
%\title{The Eruptive Coupling of Helmet Streamers and Pseudostreamers}
%\title{A New Class of Sympathetic Solar Eruptions}
%A Three Dimensional Simulation of a Pseudostreamer Coronal Mass Ejection}
%\title{How Similar are Pseudostreamer Filament Eruptions to Mini-filament jets?}


\author{P.~F.~Wyper} 
\affil{Department of Mathematical Sciences, Durham University, Durham, DH1 3LE, UK}
\email{peter.f.wyper@durham.ac.uk}

\author{S.~K.~Antiochos} 
\affil{Heliophysics Science Division, NASA Goddard Space Flight Center, 8800 Greenbelt Rd, Greenbelt, MD 20771, USA}
\email{spiro.antiochos@nasa.gov}

\author{C.~R.~DeVore} 
\affil{Heliophysics Science Division, NASA Goddard Space Flight Center, 8800 Greenbelt Rd, Greenbelt, MD 20771, USA}
\email{c.richard.devore@nasa.gov}

\author{B.~J.~Lynch} 
\affil{Space Sciences Laboratory, University of California, Berkeley, CA 94720, USA}
\email{blynch@ssl.berkeley.edu}

\author{J.~T.~Karpen} 
\affil{Heliophysics Science Division, NASA Goddard Space Flight Center, 8800 Greenbelt Rd, Greenbelt, MD 20771, USA}
\email{judith.t.karpen@nasa.gov}

\author{P.~Kumar} 
\affil{Heliophysics Science Division, NASA Goddard Space Flight Center, 8800 Greenbelt Rd, Greenbelt, MD 20771, USA}
\affil{Department of Physics, American University, Washington, DC 20016, USA}
\email{pankaj.kumar@nasa.gov}

\begin{abstract}
A highly important aspect of solar activity is the coupling between eruptions and the surrounding coronal magnetic-field topology, which determines the trajectory and morphology of the event and can even lead to 
%so-called 
{{sympathetic}} eruptions from multiple sources. In this paper, we report on a numerical simulation of a new type of coupled eruption, in which a coronal jet initiated by a large pseudostreamer filament eruption triggers a streamer-blowout coronal mass ejection (CME) from the neighboring helmet streamer. Our configuration has a large opposite-polarity region positioned between the polar coronal hole and a small equatorial coronal hole, forming a pseudostreamer flanked by the coronal holes and the helmet streamer. Further out, the pseudostreamer stalk takes the shape of an extended arc in the heliosphere. We energize the system by applying photospheric shear along a section of the polarity inversion line within the pseudostreamer. The resulting sheared-arcade filament channel develops a flux rope that eventually erupts as a classic coronal-hole-type jet. However, the enhanced breakout reconnection above the channel as the jet is launched progresses into the neighboring helmet streamer, partially launching the jet along closed helmet streamer field lines and blowing out the streamer top to produce a classic bubble-like CME. This CME is strongly deflected from the jet's initial trajectory and contains a mixture of open and closed magnetic field lines. We present the detailed dynamics of this new type of coupled eruption, its underlying mechanisms and the implications of this work for the interpretation of in-situ and remote-sensing observations. 
\end{abstract}


%%%%%%%%%%

\section{Introduction}
The solar corona hosts a tremendous amount of eruptive activity and flare energy release that plays out across a vast range of sizes and time scales. At the largest scales, eruptive active-region flares produce substantial bubble-like coronal mass ejections (CMEs) that can strongly influence near-Earth space weather \citep[e.g.][]{Webb2012}. At the smallest end of the spectrum, tiny filament-channel eruptions form coronal jets \citep[e.g.][]{Sterling2015}. The ultimate unifying feature of all these eruptions are filament channels, consisting of strongly sheared magnetic field lines that follow polarity inversion lines (PILs) of the normal magnetic field component on the solar surface \citep{Martin1998}. Filament channels provide the free magnetic energy for the eruption, but it is the interaction between the filament channel and the surrounding magnetic field that dictates how the eruption is triggered and its eventual morphology.

A prime example of this variety is the array of eruption morphologies that result from filament channels formed within multipolar magnetic topologies. At the largest scales, these eruptions can be triggered by the systematic removal of strapping flux by reconnection at a coronal null point, magnetic breakout \citep{Antiochos1999}. Such events ultimately lead to fast, large-scale, bubble-like CMEs \citep[e.g.,][]{Lynch2008,Lynch2009,Lynch2016,Karpen2012,Masson2013,Masson2019,Chen2016,Dahlin2019}. At much smaller scales, both observations \citep[e.g.][]{Sterling2015,Moore2018,Kumar2018,Kumar2019} and simulations \citep[e.g.][]{Wyper2017,Wyper2018,Wyper2019} have shown that the same eruption mechanism is at work in the mini-filament eruptions that form many coronal jets. The key difference determining the eruption morphology is how the erupting flux rope that forms from the filament channel interacts with the surrounding magnetic field. In jets, the flux rope reconnects at the null point low in the corona, transferring magnetic twist and filament plasma to the surrounding open field and creating a narrow plasma ejection that adds no new open flux to the heliosphere. In eruptions at active-region scales, on the other hand, the erupting flux rope remains connected to the surface at both ends, opening new flux into the heliosphere as part of the bubble-shaped CME. The local magnetic environment around the filament channel clearly plays a crucial role in its eventual eruptive morphology.

\begin{figure}
\centering
\includegraphics[width=0.45\textwidth]{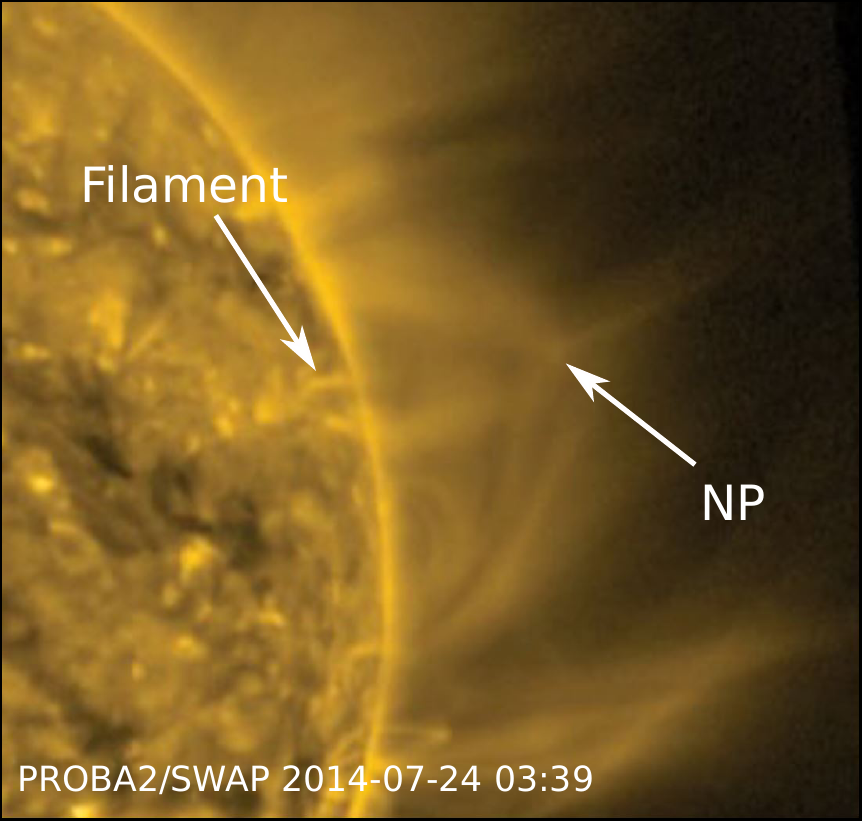}
\caption{PROBA2/SWAP 174\,\AA\, processed image of a pseudostreamer. NP = apparent null point lying between two open field regions.}
\label{fig:swap}
\end{figure}

Perhaps the most definitive, and certainly the most striking, example of the interaction between the filament channel and its surrounding is the phenomenon of sympathetic eruptions. That flares can be sympathetic has been known for many decades \citep[e.g.][]{Richardson1951}, but sympathetic eruptions have attracted much attention recently due to the high-cadence, full-Sun coronal observations of \emph{STEREO} and \emph{SDO} \citep[e.g.,][]{Schriver2011,Titov2012,Jin2016}. By examining a series of eruptions that included filament ejections, flares, and CMEs, and whose locations covered a large fraction of the solar surface, \citet{Schriver2011} presented compelling observational evidence that the sympathetic nature was due to the reconnection in the corona of neighboring magnetic flux systems, as occurs in the breakout model. Subsequent numerical modeling by \citet{Torok2011} and \citet{Lynch2013} demonstrated that null-point reconnection in the corona, as in breakout, naturally couples different flux systems and leads to the sympathetic eruptions. A key feature of these and other studies, both observational and theoretical, is that each eruption has its own filament channel. Consequently, each flux system is primed to erupt and one simply destabilizes the next. Our new work presented here, in contrast, shows that a coupled eruption of multiple flux systems can be driven by a single filament channel. Furthermore, it shows that breakout reconnection can energize, as well as destabilize, coupled eruptions.

The vast difference in scales and energies between eruptions that produce large-scale, bubble-like CMEs versus small-scale, collimated coronal jets implies that there is a continuum of eruption scenarios in between, unified by the role of breakout reconnection in dictating the eruption morphology. One scenario in this middle ground is a filament eruption from a pseudostreamer. At first glance, pseudostreamers have the same topology as coronal jets, simply on a much larger scale. Figure \ref{fig:swap} shows a SWAP image of a pseudostreamer harbouring a small filament observed on the limb. In profile, the pseudostreamer topology resembles that of a single null point above two coronal arcades, which sit between coronal holes of like polarity. The presence of this multipolar topology assures that, like mini-filament coronal-hole jets, pseudostreamers can host filament eruptions that occur via magnetic breakout.

\begin{figure*}
\centering
\includegraphics[width=\textwidth]{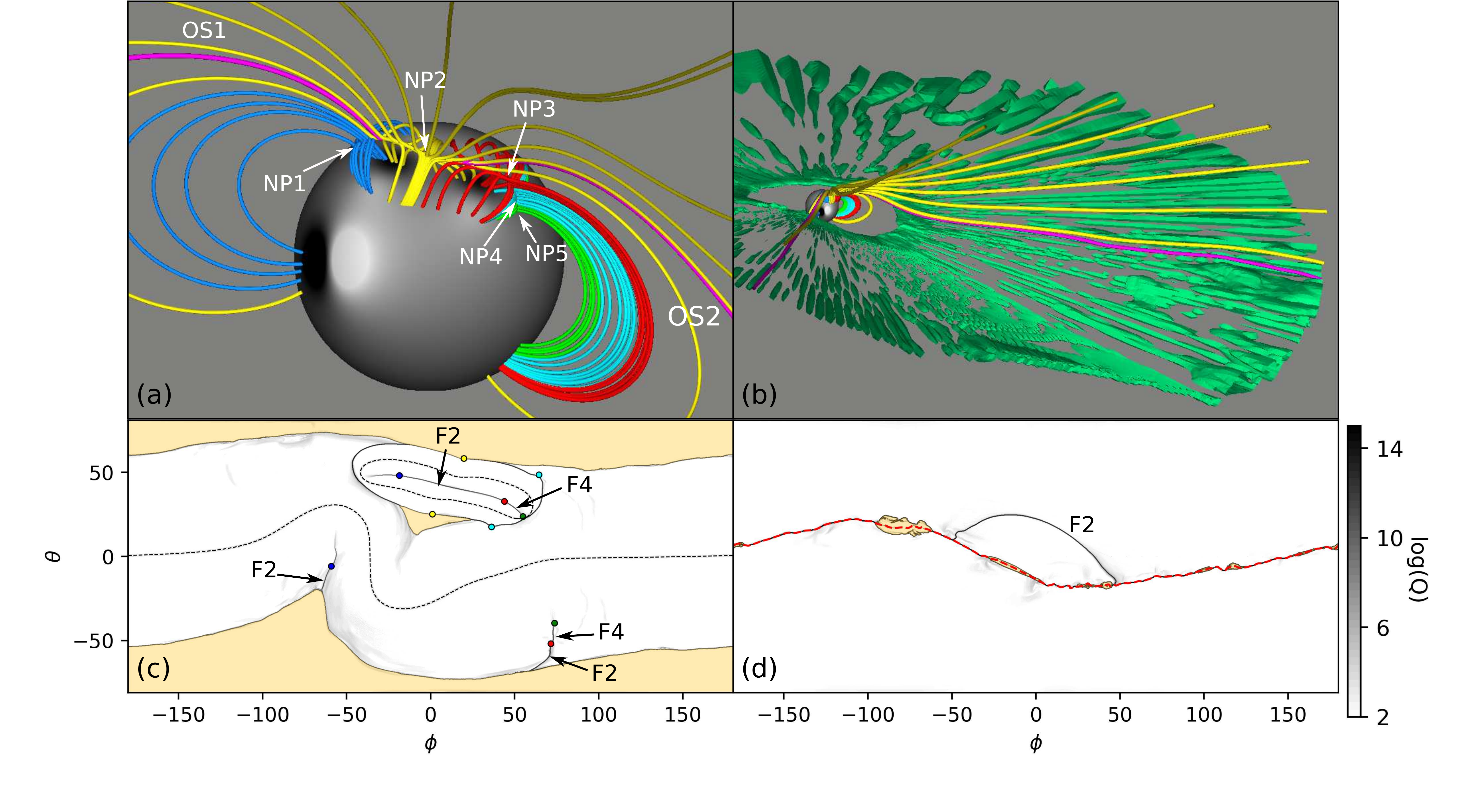}
\caption{{{Overview of the topology and connectivity of the magnetic configuration hosting the eruption.}} (a) Field lines showing the five null points (NP1, ..., NP5) that define the pseudostreamer. Two open separators (OS1 and OS2) are shown in magenta. (b) A zoomed-out view showing that the fan plane of the central null (NP2) maps to an arc in the heliosphere. Isosurfaces show regions of high plasma $\beta$ within the centre of the heliospheric current sheet. (c) $\log(Q)$ on the solar surface. PILs are shown with dashed lines. Open field regions are shaded yellow, showing the disconnected coronal hole. Circles show the spine-line footpoints of the nulls (coloured as in (a)). F2 and F4 are high-$Q$ strips associated with NP2 and NP4, respectively. (d) $\log(Q)$ at $30R_{s}$ showing the S-Web arc around the open field of the disconnected coronal hole. Yellow regions show disconnected flux. The dashed red line shows the PIL of $B_r$ within the heliospheric current sheet.}
\label{fig:relax}
\end{figure*}

In general, however, the large-scale nature of pseudostreamers leads to a much richer variety of magnetic structure than those that underlie jets. The closest to jets are small pseudostreamers associated with newly emerged active regions within low-latitude coronal holes. These pseudostreamers have a quasi-circular base, anemone-like when view from above, with a single magnetic null \citep{Shibata1994,Asai2008,Kumar2020} in the ``embedded bipole'' topology \citep{Antiochos1990}. More complex are the pseudostreamers that form above the long tails of minority polarity, stretched out by differential rotation and meridional flow, that are associated with decaying active regions. These often involve multiple nulls and/or bald patches along the length of the pseudostreamer, and they are so large that their proximity to the nearby helmet-streamer boundary also must be taken into account \citep{Titov2011,Titov2012,Scott2018,Scott2019,Masson2019}. 

Large pseudostreamers are key contributors to the complexity of the S-Web \citep{Antiochos2011}, in which the outer spines and/or fan planes of the nulls separating the magnetic flux of neighbouring coronal holes form broad arcs that extend far out into the heliosphere. Pseudostreamers bordering equatorial coronal holes and their associated S-Web arcs are common features of the corona throughout the sunspot cycle \citep{Scott2019}. Yet to date, no simulation studies that we have seen model a large-scale pseudostreamer eruption and examine its effects on the global helmet streamer and the S-Web. 

In this work, we present a simulation that addresses both of these interactions, along with the specifics of the pseudostreamer eruption itself. We constructed a large pseudostreamer that has multiple null points and separator lines {{(Fig. \ref{fig:relax}(a))}}, is bounded north and south by the polar and an equatorial coronal hole, and that partially underlies the global helmet streamer that forms the polar coronal-hole boundary to the east and west of the pseudostreamer {{(Fig. \ref{fig:relax}(c))}}. The pseudostreamer is rooted in a large, elliptical region of minority polarity that represents a decayed active region. There are several important differences between this simulation and our previous studies of jets and CMEs. First, the ``overlying'' coronal field consists of both open (coronal-hole) and closed (helmet-streamer) flux. In earlier work, the overlying field was either entirely open (jets) or entirely closed (CMEs). The separatrix dome of the minority polarity forms a section of the open/closed flux boundary and connects the polar and equatorial coronal holes. In addition, the dome connects the pseudostreamer to the neighboring helmet streamer, so that reconnection between the pseudostreamer and helmet-streamer fluxes is possible. As we shall see, the breakout reconnection indeed involves both the usual open/closed interchange reconnection of jets \citep{Wyper2018} and the usual closed/closed reconnection of CMEs \citep[e.g.,][]{Masson2019,Dahlin2019}. Second, the reconnection in the corona can occur at multiple null points and along the separator line. Consequently, it could be expected to be much more efficient than single-null-point reconnection, but the effect this should have on the eruption is not clear. Highly efficient breakout reconnection may limit the amount of free energy that can be built up in the filament channel; but it also may allow more of the stored free energy to be released. Third, as will be shown below, the filament channel that we formed extends over a rather small fraction ($< 25$\%) of the PIL of the minority polarity. For a single null point, it is not obvious that an explosive eruption could be obtained with such a small filament channel. Highly efficient, multiple null-point breakout reconnection could play a critical role in the vigor of an eruption from our pseudostreamer. 

The results presented below show that the low-coronal evolution of our pseudostreamer eruption is quite similar to that of a mini-filament jet, and that breakout reconnection plays a similarly central role in dictating the eruption morphology. However, rather than producing a collimated jet-like expulsion of mass in the outer corona, our simulated eruption generates a bubble-like CME by blowing out the top of the adjacent helmet streamer. The ejected magnetic flux is comprised of a mixture of open and closed field lines that broadly follow the S-Web arc. This result has important implications for in-situ observations, as we briefly discuss. In \S \ref{sec:model} we give the simulation details. \S\S \ref{sec:pre}, \ref{sec:low}, and \ref{sec:energy} describe, respectively, the pre-eruptive changes to the magnetic field, the low-coronal evolution of the eruption, and the energy release. \S\S \ref{sec:helmet} and \ref{sec:qmap} examine the subsequent blowing out of the nearby helmet streamer. \S\S \ref{sec:cme} and \ref{sec:sep} describe the CME and the potential associated impulsive solar energetic particle (SEP) signatures, respectively. We discuss our findings and present our conclusions in \S \ref{sec:disc}.

\begin{figure*}
\centering
\includegraphics[width=0.9\textwidth]{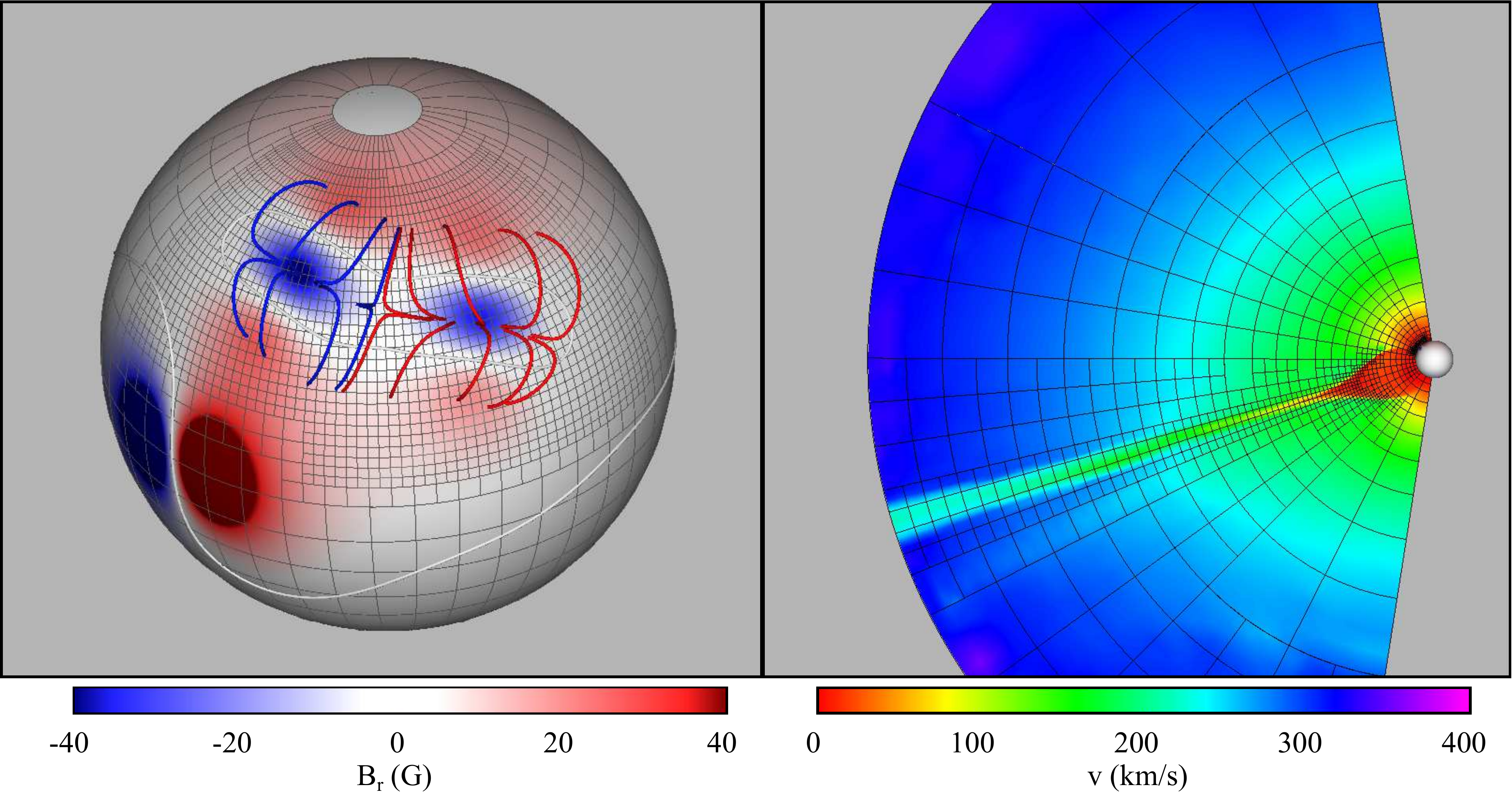}
\caption{Simulation grid (a) on the surface, with field lines from near the fan planes of null points NP1 and NP3 (see Fig.\ \ref{fig:relax}) drawn for context, and (b) in a plane of constant longitude, showing how the adaptive refinement localises the grid to the heliospheric current sheet.}
\label{fig:grid}
\end{figure*}

\section{Model Description}
\label{sec:model}
\subsection{Equations and Initial Condition}
\label{sec:setup} 
The simulation was performed using the Adaptively Refined Magnetohydrodynamics Solver (ARMS), which uses a Flux-Corrected Transport algorithm to capture shocks and minimize numerical diffusion \citep{DeVore1991}. The following ideal MHD equations were solved in spherical coordinates: 
\begin{gather}
\frac{\partial \rho}{\partial t} + \boldsymbol{\nabla}\cdot(\rho \mathbf{u}) = 0, \\
\frac{\partial \rho \mathbf{u}}{\partial t} + \boldsymbol{\nabla}\cdot(\rho \mathbf{u}\mathbf{u}) = \frac{1}{4\pi}(\boldsymbol{\nabla}\times\mathbf{B})\times\mathbf{B} - \boldsymbol{\nabla}P + \rho\mathbf{g}, \\
\frac{\partial \mathbf{B}}{\partial t} - \boldsymbol{\nabla}\times(\mathbf{u}\times\mathbf{B}) = 0,
\end{gather}
where $\rho$ is the plasma mass density, $\mathbf{u}$ the plasma velocity, $\mathbf{B}$ the magnetic field, $P$ the plasma pressure, and $\mathbf{g} = -G M_{\odot}\mathbf{r}/r^3$ the gravitational acceleration. Magnetic reconnection in the model occurs due to numerical diffusion associated with the algorithm. We assume a fully ionised hydrogen plasma, so that $P = 2(\rho/m_p)k_B T$. The temperature is further assumed to be constant and uniform throughout the volume with $T = 1$\,MK. 

The simulation volume is given by $\phi \in [-180^{\circ},180^{\circ}$] in longitude, $\theta \in [-81^{\circ},+81^{\circ}]$ in latitude, and $r\in [1R_{s},30R_{s}]$ in radius. The domain is periodic in $\phi$. At the inner radial boundary, mass is allowed to flow into, but not out of, the volume. The three radial guard cells below the inner boundary are fixed at their initial densities, with their velocity components set to zero. These cells act as a reservoir of mass sustaining the wind solution described below. The magnetic field at the radial inner boundary is line-tied, with the tangential velocity components set to zero except where prescribed otherwise by the driving flow given below. At the outer radial boundary, flow-through (zero-gradient extrapolated) conditions are applied to the radial velocity component and half-slip (zero-value outside) conditions are applied to the tangential components. On the $\theta$ boundaries, the normal velocity is reflecting and the tangential components are free-slip (zero-gradient). Altogether, these boundary conditions maintain a quasi-steady, isothermal solar wind throughout the open field in the domain.

The initial magnetic field is given by a potential-field source-surface (PFSS) solution, with the source surface at $3R_{s}$. The magnetic field at the radial inner boundary is defined analytically by combining sub-surface magnetic dipoles with a Sun-centered dipole that defines the global dipolar magnetic field. The global dipole is chosen so that $|B_r| = 10$\,G at the poles. The sub-surface dipoles are placed so as to create a strip of negative polarity in the northern hemisphere, bordering a strong equatorial bipolar active region, Fig.\ \ref{fig:grid}(a). The equatorial active region acts to pull the helmet streamer southward whilst the strip of negative polarity cuts off a section of northern polar coronal hole, forming a disconnected coronal hole separated from the north-pole hole by a pseudostreamer \citep{Titov2011,Scott2018} described further below. Such a strip of opposite-polarity flux is a common feature in magnetograms, typically forming as active regions decay and their trailing polarity ``rushes'' to the pole due to meridional flow \citep[e.g.][]{Howard1981,Wang1989}.

The atmosphere was initialized by a 1D \citet{Parker1958} wind solution of the form 
\begin{equation}
\frac{v^2}{c_s^2}\exp\left(1-\frac{v^2}{c_s^2}\right) = \frac{r_s^4}{r^4}\exp\left(4-4\frac{r_s}{r}\right).
\end{equation}
Here $v(r)$ is the radial velocity, $c_s = \left( 2k_BT_0/m_p \right)^{\frac{1}{2}}$ is the isothermal sound speed, and $r_s = GM_\odot m_p/4k_BT_0$ is the radius of the sonic point. With $T_0=1\times 10^6$\,K, $v = c_s = 128$\,km/s at $r = r_s = 5.8 R_{\odot}$. The plasma number density at the base of the atmosphere is a free parameter that we set to $7.2\times 10^9$\,cm$^{-3}$. This value gives a minimum plasma $\beta$ (ratio of thermal to magnetic pressure) in the vicinity of the pseudostreamer of about 10\%{{, so the dynamics is properly field-dominated. The density and the resultant Alfv\'en speed are intended to be more typical of closed-field regions below pseudostreamers and helmet streamers, where our eruption occurs, than of open-field regions in the neighboring coronal holes}}.

The magnetic field defined by the PFSS solution and the plasma wind solution are not initially in equilibrium, so before the surface driving was applied the simulation was first run out through a long relaxation phase ($> 1\times 10^6$\,s) until the plasma and magnetic field evolved to near pressure balance. Figure \ref{fig:grid}(b) shows the plasma velocity in a plane of constant longitude after the relaxation. Away from the heliospheric current sheet, the solar wind reaches $\approx 350$\,km/s at $30R_{s}$. This wind speed, more typical of slow rather than fast wind, is due to the low temperature $T$ and our simplifying isothermal approximation. However, our main goals of tracking with high fidelity the low-coronal evolution of the eruption and its interaction with the helmet streamer are not affected significantly by the asymptotic wind speed. 

The base grid is uniformly spaced in $\phi$ and $\theta$ and is stretched exponentially in radius. Four levels of grid refinement were allowed in the simulation. A fixed region of maximal resolution was included that encompassed the pseudostreamer. Figure \ref{fig:grid}(a) shows the grid blocks on the surface in this region. Each grid block consists of $8\times8\times8$ grid cells. The minimum angular grid spacing is $\approx 0.28^{\circ}$ in both $\phi$ and $\theta$, corresponding to a maximum grid resolution (at the finest grid level) on the inner radial boundary of $\approx 3.4$\,Mm. In the rest of the volume, the grid was adapted dynamically based on local gradients in the magnetic field (see \citet{Karpen2012} for details of the method). The parameters were chosen to maximally resolve the heliospheric current sheet (Fig.\ \ref{fig:grid}(b)), as well as the shocks and current sheets associated with the CME once underway. To make the calculation more tractable and reduce the overall grid size, the maximal resolution on the back side of the Sun ($\phi \in [-180,-90] \cup [+90,+180]$) was limited to three, rather than four, levels of refinement.

\subsection{Topology of the Relaxed State}
The magnetic topology of the pseudostreamer after the relaxation is shown in Figure \ref{fig:relax}(a). The surface $B_r$ distribution was chosen such that there were two local minima of $B_r$ within the negative-polarity strip, Figure \ref{fig:grid}(a). This naturally creates a system of at least three nulls, one associated with each minimum (NP1 and NP3) and another that resides between them (NP2). Two separators connect the three nulls along the top of the separatrix surface dome. In the PFSS solution only these three nulls were present. However, during the relaxation currents naturally developed around the nulls and along the separators. Combined with changes in the field around the pseudostreamer as the helmet streamer relaxed, this led to a bifurcation of NP3 and the formation of two further nulls, labeled NP4 and NP5, connected by an additional two separators. Due to where we store the magnetic free energy in our simulation, these additional nulls are not involved in the eruption. 

The key elements of the pseudostreamer topology can be understood by considering NP1, NP2, and NP3 (with NP4 and NP5 understood by extension). The fan planes of NP1 and NP3 form the main sections of the separatrix surface that divide the closed field beneath the pseudostreamer from the surrounding open (or distantly closing) field in the rest of the corona (blue and red field lines, Fig.\ \ref{fig:relax}(a)). The inner spines of each null connect to the strip of negative $B_r$ beneath the pseudostreamer; the outer spines connect to negative $B_r$ in the southern hemisphere. Both nulls therefore reside in the closed field beneath the helmet streamer.

NP2 sits at the intersection of the fan planes of NP1 and NP3. The spines and fan of NP2 are oppositely oriented to those of NP1 and NP3, so that the spines of NP2 sit on the separatrix surface of the pseudostreamer and its fan plane is aligned radially (yellow field lines, Fig.\ \ref{fig:relax}(a)). The spines of adjacent nulls actually bound the fan plane of the other, so the radial partially open fan plane of NP2 is bounded on either side by the closed spines of NP1 and NP3. As a consequence, the fan of NP2 must straddle the open-closed separatrix of the helmet streamer (note the closed yellow field lines adjacent to NP1 and NP3, Fig.\ \ref{fig:relax}(a)). Therefore, there must be two separators connecting the base of the heliospheric current sheet with NP2 (shown in magenta, Fig.\ \ref{fig:relax}(a)). A similar alternation of null orientation occurs for NP3, NP4, and NP5, although the fan plane of NP4 lies entirely within the closed field beneath the helmet streamer. 

The surface connectivity of the relaxed state is shown in Figure \ref{fig:relax}(c). Yellow regions show open field, highlighting the triangular low-latitude corona hole cut off by the pseudostreamer. The logarithm of the squashing factor \citep[$Q$;][]{Titov2007} is shown in gray scale. Strips of high $Q$ show regions where the field-line connectivity varies rapidly between adjacent footpoints, and $Q$ is formally infinite at separatrix surfaces. The footprint of the global helmet streamer is evident, along with a closed curve of high $Q$ showing the footprint of the pseudostreamer separatrix dome. The footpoints of the spines of each null and the strips of high $Q$ associated with fan planes of NP2 and NP4 are also highlighted (F2 and F4, respectively). The fan plane of NP2 is truly global, in that it connects to the surface beneath the pseudostreamer, to distant positions in the southern hemisphere, and also out into the solar wind, where it forms an S-web arc \citep{Antiochos2011} dividing the open fluxes of the equatorial and northern polar coronal holes. Field lines within the arc and the associated high-$Q$ arc at $30R_{s}$ are shown in Figure \ref{fig:relax}(b) and (d). Given such global connectivity, one should expect a global influence of the eruption, which indeed is what we find. Figure \ref{fig:relax}(d) also shows several small regions of disconnected magnetic flux that are associated with {{concave-up field lines (``U-loops'') that enter and leave the domain through the outer radial boundary and localised bundles of spiral field lines and compressed plasma (``plasmoids'')}} within the heliospheric current sheet. These features are formed and expelled periodically as part of the dynamic quasi-steady state of the magnetic structure \citep{Higginson2017,Titov2017}.

\begin{figure*}
\centering
\includegraphics[width=\textwidth]{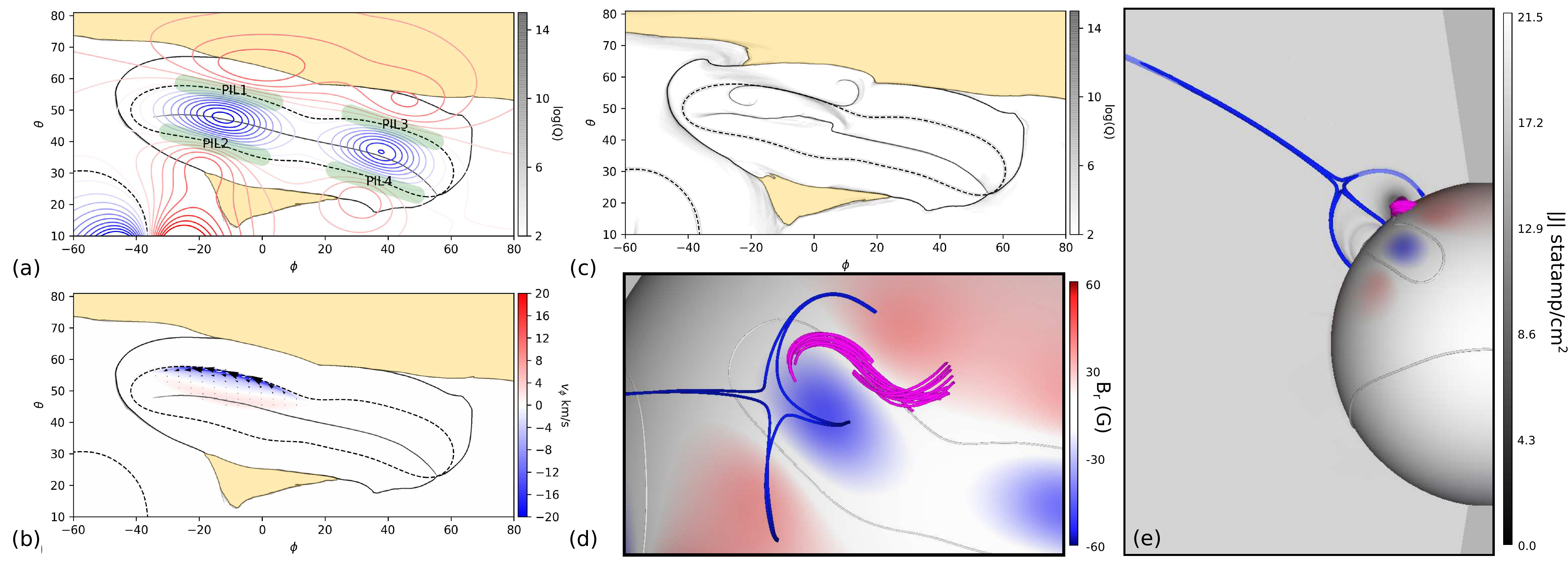}
\caption{(a) Log $Q$ and coronal holes overlaid with contours of $B_r$ (between $-40$\,G and $40$\,G); the four sections of PIL bordered by strong concentrations of $B_r$ are shown in green. (b) Arrows show the direction and strength of the shear driving; red/blue shading shows the magnitude of the $v_{\phi}$ component. (c) Log $Q$ at $t = 15$\,hrs $50$\,min showing two $J$-shaped hooks characteristic of a sigmoidal field. (d) Field lines in the filament channel showing its sigmoidal shape. (e) Side view showing the inflation of the pseudostreamer and the opening of the outer spine of NP1.}
\label{fig:filpos}
\end{figure*}
%2221878

\begin{figure}
\centering
\includegraphics[width=0.5\textwidth]{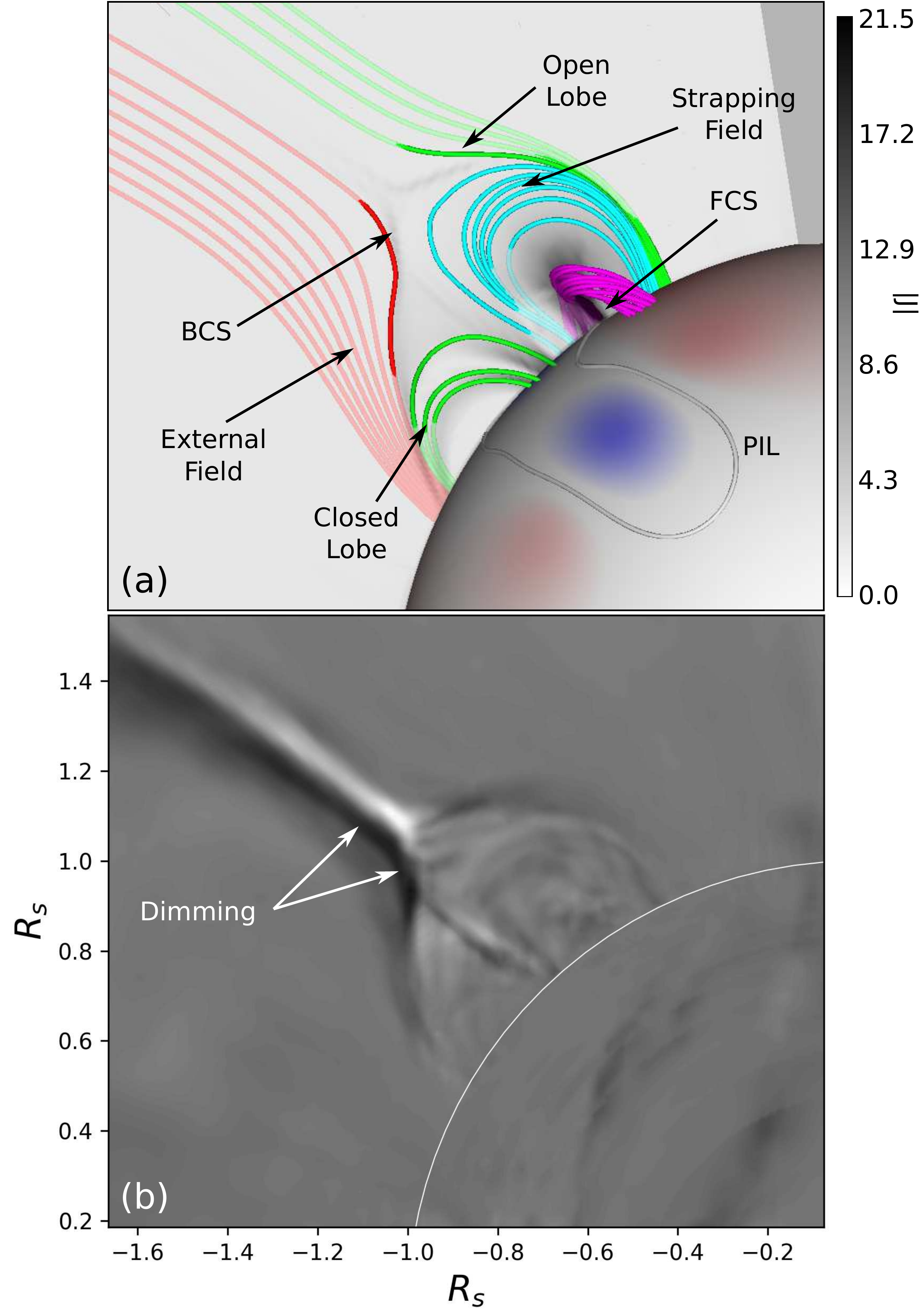}
\caption{(a) Pre-eruption magnetic field at $t = 24$\,hrs $10$\,min. BCS = breakout current sheet, FCS = flare current sheet. Field lines indicate the four flux regions separated by the BCS. {{The semi-transparent grey shading shows current density (statamp cm$^{-2}$).}} (b) Synthetic EUV base-difference image (from $t = 22$\,hrs $30$\,min) showing the density depletion above the BCS. }
\label{fig:pre}
\end{figure}
%2280912

\subsection{Filament Channel Formation}
\label{sec:driving}
In the classic 2D picture of a pseudostreamer, filament channels can form at the PILs beneath either or both of the two closed field regions associated with the null point. In 3D, these two separate PILs join to form a closed loop. Theoretically, filament channels can form along any section of this PIL, but the most energetic would be expected to form between strong concentrations of $B_r$ associated with stronger strapping field. That is, these sections of PIL are able to support {{higher energy density (higher magnetic pressure)}} because of their increased strapping field. Due to the distribution of $B_r$ on the solar surface, chosen so as to generate the pseudostreamer topology described above, there are four sections of PIL bordered by local peaks in $|B_r|$ as shown in green in Figure \ref{fig:filpos}(a). These various locations for energetic filament-channel formation would be expected to stress different nulls and separators in the pseudostreamer topology, with associated differences in the connectivity changes occurring during eruption. Furthermore, combinations of filament channels in different locations can produce sympathetic eruptions \citep[e.g.][]{Torok2011,Lynch2013}. 

We performed several simulations in which filament channels formed along each of the highlighted sections. In this paper, we focus on the case where a filament channel is formed at PIL1, leaving the exploration of eruption morphologies and coupling behaviour of filament channels formed along the other sections to future studies. To create the filament channel, a tangential velocity field was imposed on the lower boundary. The velocity profile, a generalisation of one used by \citet{Higginson2017}, is an elliptical flow pattern, tilted with respect to the $\theta,\phi$ coordinates, that preserves $B_r$ on the surface throughout the evolution. {{It is shown in Figure \ref{fig:filpos}(b).}} In order to satisfy 
\begin{equation}
\frac{\partial B_r}{\partial t} = -\boldsymbol{\nabla}_{\perp}\cdot(\mathbf{v}_{\perp}B_r) = 0,
\end{equation}
$\mathbf{v}_{\perp}$ is constructed from the curl of a radial vector,
\begin{equation}
\mathbf{v}_{\perp} = \frac{1}{B_r} \boldsymbol{\nabla}_{\perp} \times (\psi,0,0),
\end{equation}
where $\psi$ is a function of $\theta$, $\phi$, $t$, and $B_r$ in the form 
\begin{equation}
\psi(\theta,\phi,t) = V_0 f(t) g(\xi) h(\eta).
\end{equation}
{{By construction, streamlines of the driving flow follow the contours of $\psi$ over a compact surface region where $\psi \ne 0$. The function $g(\xi)$ defines a simple tilted ellipse that depends upon the spatial coordinates $(\theta,\phi)$; $h(\eta)$ depends solely upon $B_r$ and serves to keep the boundary of the flow region slightly removed from the PIL, so the latter is not distorted by the flow.}} 
We choose 
\begin{gather}
g(\xi) = \frac{m+\ell+1}{\ell+1} \left[ 1 - \xi^{2(\ell+1)} \right] - \left[ 1 - \xi^{2(m+\ell+1)} \right],\\
h(\eta) = \eta,
\end{gather}
where we set $m = \ell = 1$ and 
\begin{gather}
\xi^2 = \min \left( \frac{\alpha^2}{a^2} + \frac{\beta^2}{b^2}, 1 \right),\\
\eta = \max \left( \min \left( B_r,B_2 \right), B_1 \right) - B_2.
\end{gather}
We define the tilted orthogonal angle coordinates 
\begin{gather}
\alpha \equiv \delta \left( \theta - \theta_0 \right) + \epsilon \left( \phi - \phi_0 \right),\\
\beta \equiv \delta \left( \phi - \phi_0 \right) - \epsilon \left( \theta - \theta_0 \right),
\end{gather}
whose direction cosines are $\delta$ and $\epsilon$, whose origin is positioned at $\left( \theta_0, \phi_0 \right)$, and whose maximum extents are $\pm a$ and $\pm b$, respectively. {{The direction cosines define the tilt angle $\arctan (\delta/\epsilon)$ of the ellipse with respect to lines of longitude; the ellipse's semi-major and -minor axes are $a$ and $b$, respectively. Within the ellipse, the flow is further restricted to the region where $B_r < B_2$, whence $h(\eta) \ne 0$.}}
We chose constants $B_1 = -50, B_2 = - 1$, $\theta_0 = +0.194\pi$, $\phi_0  = -0.056\pi$, $\delta = +0.195, \epsilon = +0.981$, and $a = 0.333\pi, b = 0.055\pi$. {{These choices centered the ellipse at $55^\circ$ latitude and $-11^\circ$ longitude, and tilted it $11^\circ$ with respect to lines of longitude, as can be seen in the figure.}}
We set the magnitude and sign of $V_0$ ($= -2.5\times 10^{15}$) to quasi-statically create a filament channel of dextral chirality, as is typically observed in the northern hemisphere \citep[e.g.][]{Martin1998}.

The driving speed peaks at about 30\,km s$^{-1}$ in a strip along the PIL, but drops to $\approx 2$\,km s$^{-1}$ in the return-flow region away from the PIL. The peak speed is highly sub-Alfv\'{e}nic and subsonic, so free energy is injected quasi-statically. The time profile for $f(t)$ is chosen so that the driving is ramped up to its maximum speed over $500$\,s, held constant for a time, and then is ramped down to zero over another $500$\,s once the eruption is underway. For convenience, henceforth we will define $t=0$ as corresponding to the start of the driving. Relative to this, the driving is stopped at $t=98,000$\,s ($27$\,hrs $13$\,min). 

Figure \ref{fig:filpos}(d) and (e) show field lines in the filament channel $15$\,hrs $50$\,min into the driving. The magenta field lines in the filament channel form a classic sigmoid shape. Blue field lines show the approximate position {{of the}} spines of NP1, the outer spine of which has opened by this time (see \S \ref{sec:qmap}). It also shows the slightly inflated shape of the separatrix. At this time, NP1 sits at $1.29$\,R$_s$ (203\,Mm above the surface), although it rises further as the filament channel continues to form. Figure \ref{fig:filpos}(c) shows that two hooked, J-shaped, high-$Q$ lines form in association with the sigmoidal field lines \citep[e.g.][]{Demoulin1996,Janvier2013}. It also shows a slight shift in the pseudostreamer separatrix dome footprint, indicating that a small amount of strapping field has reconnected, along with a retreat of the adjacent helmet-streamer boundary, which now has a snub-nosed rather than tapered appearance. {{The snub-nosed shape is a signature that the helmet streamer is no longer connected to NP2 on this side. Moreover, an extremely thin corridor of open flux has formed to connect the equatorial and polar coronal holes: the widening of the thin corridor at its ends, where the open flux of the corridor meets the open flux of the coronal hole, is responsible for the shape of the boundary. Corridor formation also is consistent with the observed shift of the outer spine of NP1 into the open field. A similar switch from tapered to snub-nosed shape can be seen in the helmet-streamer boundary footprint of \citet{Titov2011} (see their Figures 5 and 6), although not explicitly noted by them in text, when a singularly thin corridor formed during an identical topological evolution.}}

\begin{figure*}
\centering
\includegraphics[width=\textwidth]{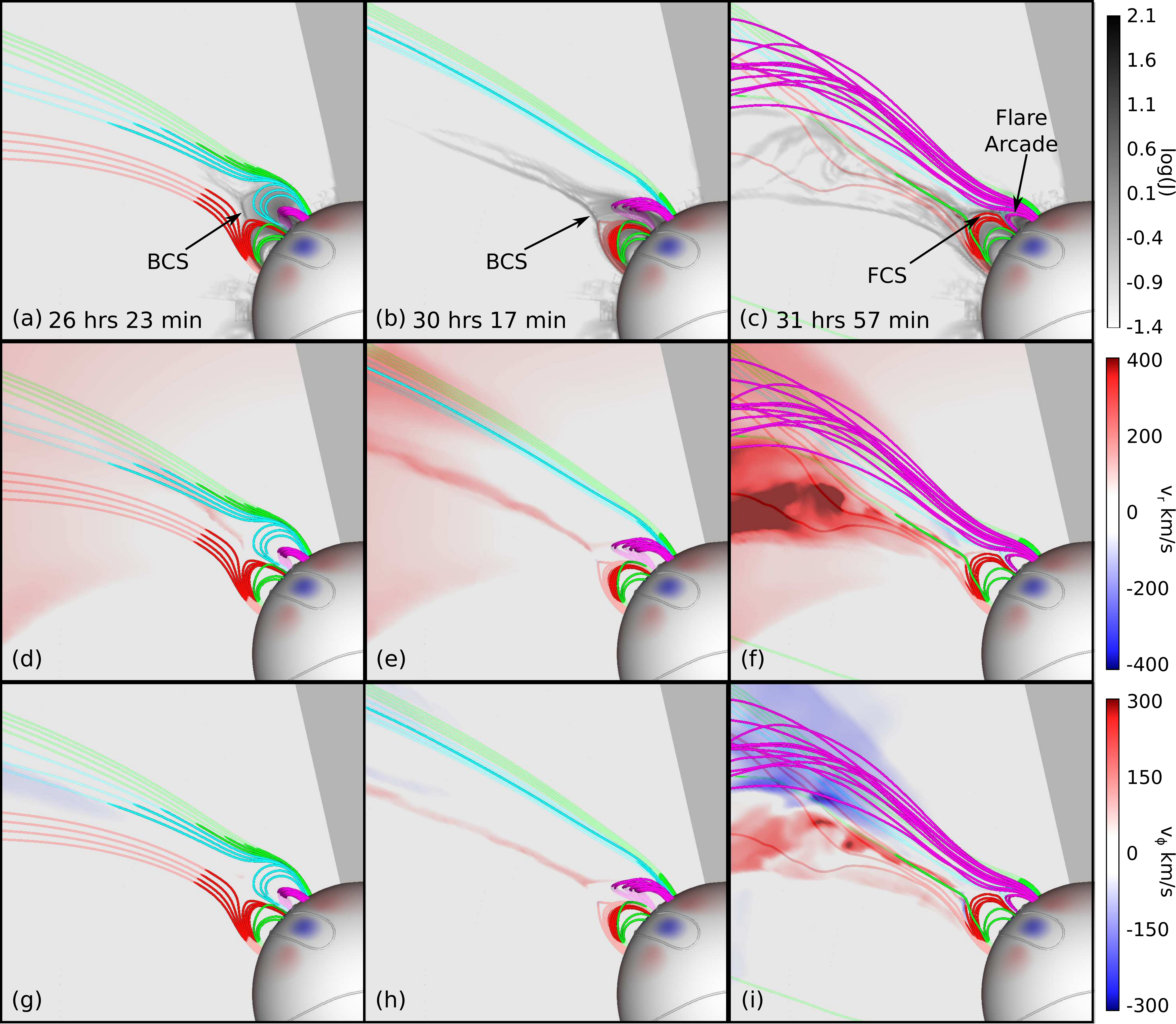}
\caption{Low-coronal eruption evolution. (a)-(c) {{Log of the current-density magnitude (statamp cm$^{-2}$)}}. (d)-(f) Radial velocity, $v_r$. (g)-(i) Line-of-sight velocity, $v_\phi$. BCS = breakout current sheet, FCS = flare current sheet. An animation of panels (a) to (c) is available online {{showing the evolution. The animation duration is 6 s.}}}
\label{fig:breakout}
\end{figure*}

\section{Pre-eruption Changes}
\label{sec:pre}
Throughout the slow driving phase prior to eruption initiation, the magnetic shear continually increases within the filament channel. The strong gradient in the driving profile adjacent to the PIL forms a concentrated volumetric current distribution along the PIL within the filament channel. Slow tether-cutting (or slipping) reconnection \citep{Moore2001,Aulanier2012} inside this current distribution gradually converts some of the sigmoid-shaped, sheared-arcade field lines into a small flux rope suspended above the PIL. For simplicity, we refer to this current concentration as the flare current sheet, although the flare reconnection begins much later in the evolution, during the explosive eruption and occurs in the lower corona below the growing flux rope. We emphasize that, at this early stage, the evolution is quasi-static and the current concentration forms entirely due to the driving profile. The flux rope begins to form about $21$\,hrs into the driving. 

Concurrent with the formation of the flux rope is the continued expansion of the strapping field above the filament channel. This expansion stresses the pseudostreamer topology, lengthening the weak current sheet already present around NP1 following the relaxation. About $t = 23$\,hrs $30$\,min into this gradual phase, systematic reconnection of the strapping field begins and feeds back upon the expansion of the filament channel, allowing the channel and embedded flux rope to rise faster. We designate this time as the onset of the breakout reconnection phase, and we denote the current sheet around NP1 as the breakout current sheet from this point onward. 

Figure \ref{fig:pre}(a) shows relevant field lines within and around the filament channel and the two current sheets at $24$\,hrs $10$\,min into the driving, shortly after breakout reconnection begins. The open lobe and the external field lines reside in the polar and equatorial coronal holes, respectively. The volumetric current distribution within the filament channel shows that the formation of the flux rope has transformed the channel flux surfaces from semi-circular to inverse-teardrop shaped. Meanwhile, the strengthening of the overlying breakout current sheet creates the characteristic cusp shape at the top of the closed-lobe region \citep[e.g.][]{Kumar2020,Wang2018}. 

Recently, \citet{Kumar2020} identified observed EUV dimmings above the breakout current sheet as a new diagnostic signaling the onset of breakout reconnection. The dimming occurs due to depletion of plasma in the surrounding volume as external field sweeps into the breakout sheet, reconnects, and flows out along the sheet as reconnection exhaust. Figure \ref{fig:pre}(b) shows a synthetic EUV base-difference image of our simulation at this time, revealing this dimming as well as the enhanced density in the reconnection outflow along the pseudostreamer spire. The subtracted base image is taken from time $t = 22$\,hrs $30$\,min, about an hour before breakout-reconnection onset. 

The presence of the flux rope makes it difficult to be unambiguous about the precise mechanism that triggers the continued rise of the flux rope towards eruption from this time onward. Breakout reconnection certainly is key to maintaining the eruption once the flux rope is rising rapidly. However, it is possible that an ideal rise of the flux rope due to torus instability couples to the breakout reconnection, to tip the evolution into a self-sustaining feedback. A similar coupling associated with ideal kink instability recently was shown to occur in simulations of active-region periphery jets \citep{Wyper2019}. Comparisons of our results with those from a perfectly ideal-MHD model \citep[e.g.][]{Pariat2009,Rachmeler2010}, which is well beyond the scope of the current investigation, would be required to address this issue definitively. However, it is clear from the fast evolution following onset of reconnection of the flux rope and from the energy plots (both described in detail below) that ideal instability of the flux rope is not the dominant process once the eruption is underway.

\begin{figure}
\centering
\includegraphics[width=0.5\textwidth]{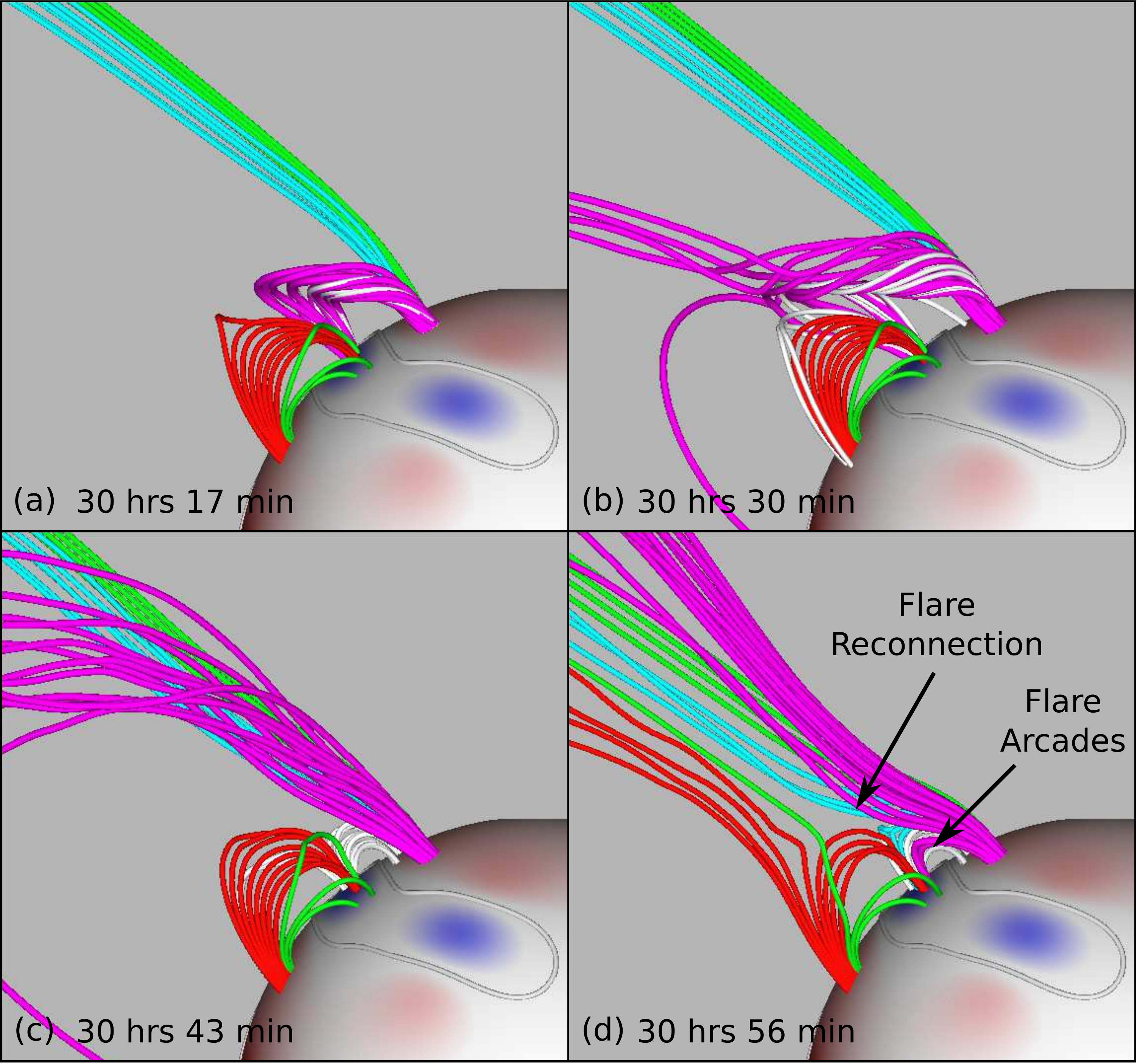}
\caption{Reconnection of the flux rope. See text {{in \S \ref{sec:rope}}} for details.}
\label{fig:rope}
\end{figure}

\begin{figure}
\centering
\includegraphics[width=0.5\textwidth]{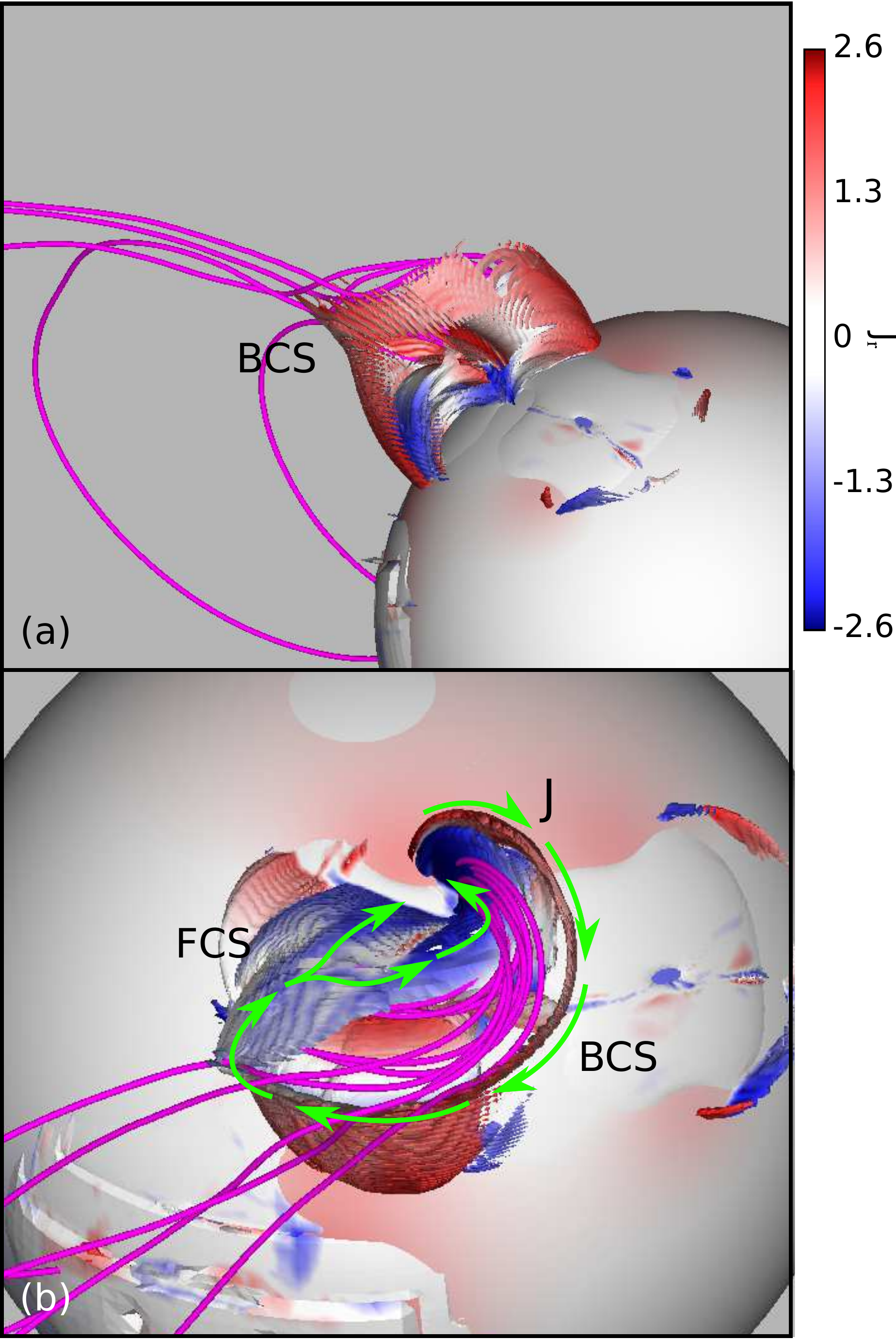}
\caption{Isosurface of current-density magnitude {{($J = 2.6$ statamp cm$^{-2}$)}}, shaded by the radial component of the current vector ($J_r$), at $t = 30$\,hrs $30$\,min. BCS = breakout current sheet, FCS = flare current sheet. (a) Side view. (b) Top view. Green arrows show the direction of the current.}
\label{fig:isoJ}
\end{figure}

\section{Low-Coronal Evolution}
\label{sec:low}
\subsection{Eruption Kinematics}
Initially, the pseudostreamer eruption closely follows the evolution of a mini-filament jet, but on a much larger scale. Once the breakout reconnection gets underway, the strapping field above the flux rope begins to be removed and the flux rope continues its rise, slowly at first. Figure \ref{fig:breakout}(a) shows the same field lines from Figure \ref{fig:pre} $2$\,hrs $13$\,min later. The field lines are traced from fixed (un-driven) points on the surface. The closing down of the red external field lines and the opening up of the cyan strapping field lines show the progression of the breakout reconnection. 

As the reconnection continues, the breakout current sheet moves southward as the external field (red field lines) sequentially reconnects. This orients the current sheet radially and positions it directly above the now-expanded (with reconnected flux) south lobe of the pseudostreamer, Figure \ref{fig:breakout}(b). As the last of the strapping flux is exhausted, the rising flux rope becomes bent towards the breakout current sheet over the top of the south lobe, Figure \ref{fig:breakout}(b). This bending of the filament-channel field, with or without filament/prominence material, is a typical feature of pseudostreamer eruptions \citep[e.g.][]{Panasenco2011,Kumar2018,Kumar2020}. In addition to bending, the flux rope accumulates more flux, as the reconnection in the flare current sheet is enhanced by the stretching of the sheet as the flux rope rises. 

When the flux rope reaches the breakout current sheet, the system departs from its previously quasi-2D evolution and enters a fully 3D phase, described below, in which the flux rope reconnects and a large fraction of the twist within the flux rope is liberated. Exactly analogous to mini-filament jets, this transfer of twist launches an untwisting plasma jet characterised by a fast radial velocity component (Fig.\ \ref{fig:breakout}(f)) and oppositely directed line-of-sight velocity components on either side of, and extended along, the jet axis; see Figure \ref{fig:breakout}(i). This shows that the low-coronal evolution of our simulated eruption is essentially that of a large-scale coronal jet.

\subsection{Flux-Rope Reconnection}
\label{sec:rope}
The reconnection of the flux rope is inherently three-dimensional, and cannot be described adequately by two-dimensional models. Figure \ref{fig:rope} shows field lines in the flux rope before, during, and after the reconnection. Magenta field lines are traced from positive-polarity footpoints, whilst silver field lines are traced from the conjugate footpoints in the negative polarity. As the flux rope begins to reconnect, some of the magenta field lines open up whilst their counterpart silver field lines become part of the southern lobe of the pseudostreamer, Figure \ref{fig:rope}(b). Figure \ref{fig:isoJ} shows the current density magnitude at this time, shaded by the radial component of the current density vector. At this crucial moment, the breakout current sheet curving around the outer edge of the flux rope and the flare current sheet following behind it combine, forming one long current sheet. The direction of the current is indicated in Figure \ref{fig:isoJ}(b) by the green arrows. This marks the instant when the null point within the breakout layer (NP1) begins to slide around the separatrix dome into the flare current sheet behind the flux rope (\S \ref{sec:qmap}). It is exactly the same topology change as occurs in our mini-filament jets \citep{Wyper2018}.

Once the null has shifted to the flare current sheet, the flare reconnection transitions from quasi-separatrix layer reconnection to null-point reconnection, and the positive-polarity footpoint of the flux rope opens completely, Figure \ref{fig:rope}(c). The breakout current sheet is carried upwards with the eruption, forming the interface between the outwardly propagating twisted field and the untwisted ambient field surrounding it. The over-pressure of the magnetic flux built up in the southern lobe during the breakout phase drives fast flare reconnection at the null, opening flux from the southern lobe whilst closing new flux into the northern lobe, the flare arcades. The flare reconnection acts to reverse much of the flux transfer that occurred during the breakout phase, re-opening external field lines (red) and re-closing strapping field lines (cyan); see Figure \ref{fig:rope}(d). Importantly for the connectivity of the CME, some of the field lines (magenta) from the positive-polarity footpoint of the flux rope get swept into the flare current sheet and also re-close, becoming part of the flare loops. That is, they become disconnected from the CME (\S \ref{sec:helmet}).

\begin{figure}
\centering
\includegraphics[width=0.5\textwidth]{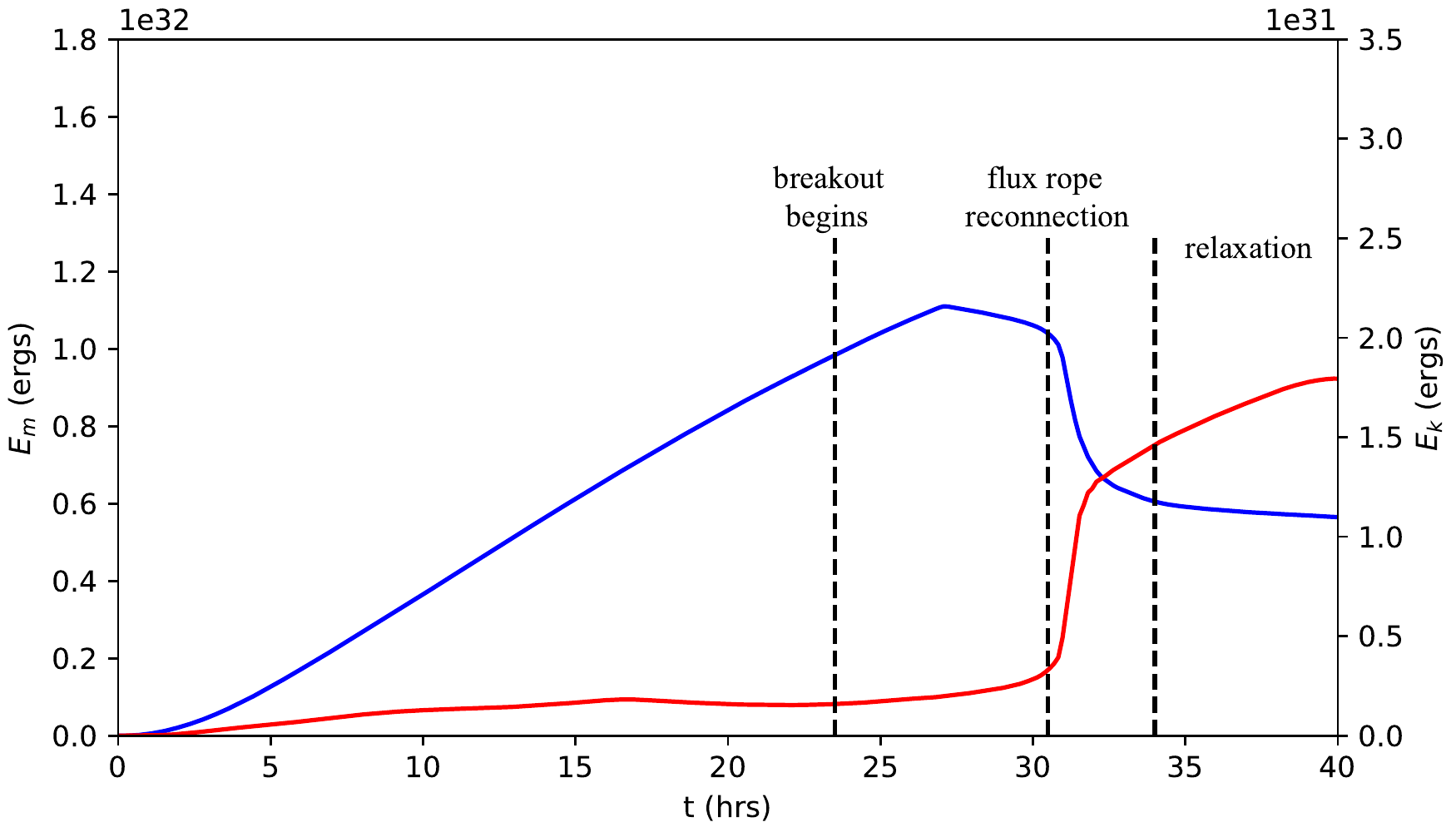}
\caption{Energy plots showing $E_m$ (blue) and $E_k$ (red) with the different eruption stages annotated.}
\label{fig:energies}
\end{figure}

\begin{figure*}
\centering
\includegraphics[width=\textwidth]{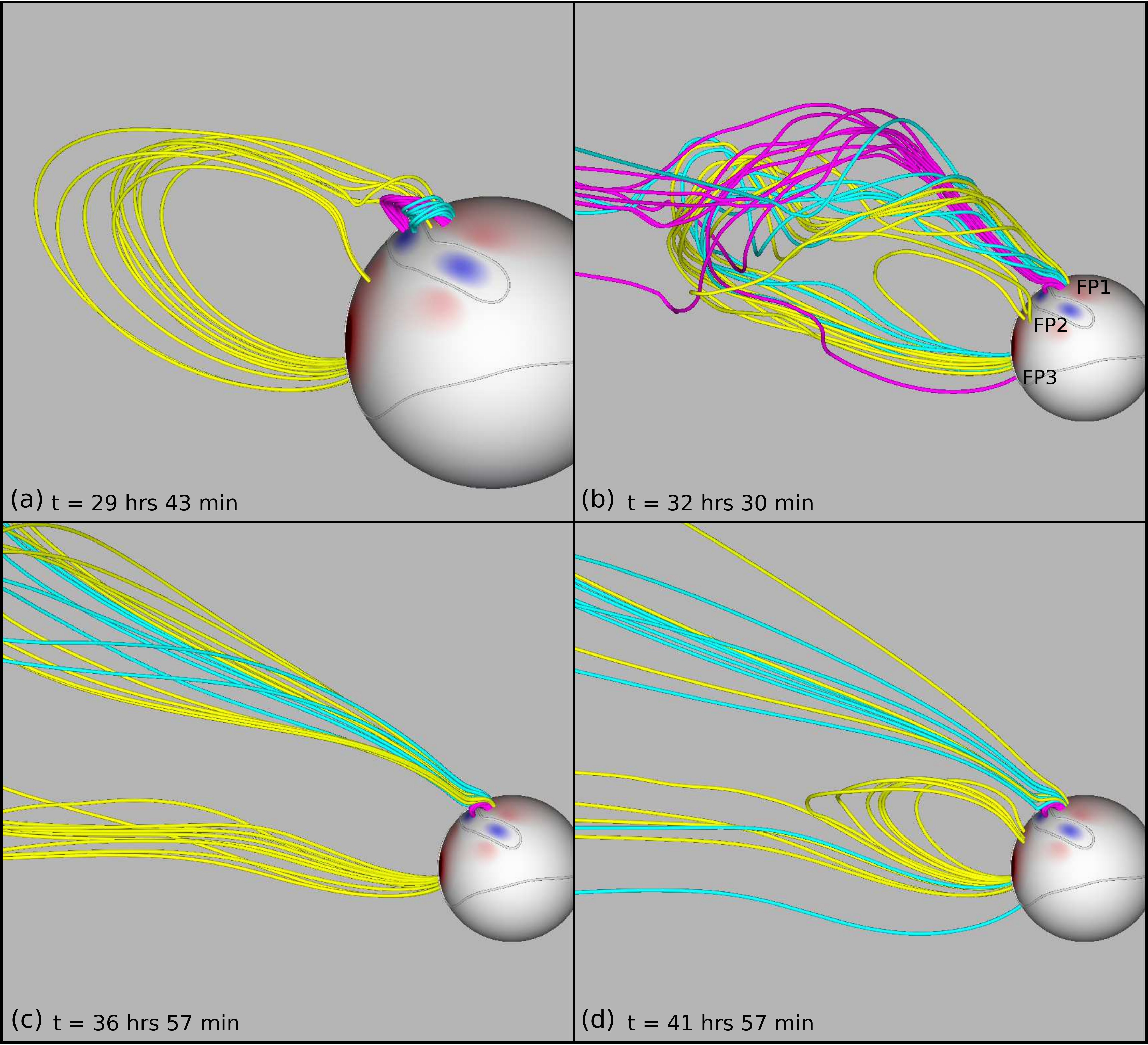}
\caption{Blowout of the helmet streamer. An animation of this figure is available online {{showing the evolution. The animation duration is 6 s.}}}
\label{fig:helmet}
\end{figure*}

\section{Energetics}
\label{sec:energy}
The evolution of the total kinetic and magnetic energies of the system further confirm the jet-like nature of the eruption. We define the free magnetic and kinetic energies as
\begin{align}
E_k &= \iiint_{V} \frac{1}{2}\rho u^2 \,dV - \left(\iiint_{V} \frac{1}{2}\rho u^2 \,dV\right)_{t=0}, \\
E_m &= \iiint_{V} \frac{1}{8\pi} B^2 \,dV - \left( \iiint_{V} \frac{1}{8\pi} B^2 \,dV\right)_{t=0},
\end{align}
where $t=0$ corresponds to the start of the driving. Although some residual fluctuations occur in the relaxed state, they are small compared to the energy stored and released by the eruption. Therefore, the above quantities provide a reasonable approximation to the actual magnetic energy stored and kinetic energy liberated by the eruption. Both curves are shown in Figure \ref{fig:energies}. The timing of the different phases of the eruption are marked with dashed lines. The kink in $E_m$ shortly after $t = 27$\,hrs coincides with the time the driving is stopped. Once the breakout reconnection is initiated ($t \approx 23$\,hrs $30$\,min), there is a gradual rise in kinetic energy, accompanied by a gradual reduction in the magnetic energy after the driving ceases. However, the major magnetic energy release occurs when the flux rope reaches the breakout current sheet and reconnects at $t\approx 30$\,hrs $30$\,min, whereupon the kinetic energy rapidly increases as the eruption is launched. The rapid changes in both energies are short-lived and slow down after about an hour, thereafter entering a more gradual relaxation phase. We found an almost identical qualitative behaviour in our previous jet simulations \citep{Wyper2017,Wyper2018}, although the kinetic energy continues to rise once the CME is launched in the present case with solar wind.

\section{Coupled Helmet-Streamer Blowout}
\label{sec:helmet}
We have seen that the low-coronal behaviour of the pseudostreamer eruption resembles that of a large-scale jet; however, the dynamics of the global topology results in a clear CME, which we now show is bubble-like. The pseudostreamer is in sufficiently close proximity to the open-closed separatrix of the helmet streamer that the helmet streamer participates in the ejection of the flux rope, as discussed further in \S \ref{sec:qmap}. Consequently, rather than transferring twist entirely onto open field lines as in a simple coronal-hole jet, a significant portion is injected into the closed field beneath the adjacent helmet streamer. This twist blows out the top of the streamer when it reaches the streamer apex. 

Figure \ref{fig:helmet}(a) shows three representative sets of field lines prior to the eruption. The positive footpoint of the erupting flux rope is shown in magenta, as in Figure \ref{fig:breakout}. A bundle of cyan field lines that initially form part of the strapping field, traced from positive footpoints slightly north of the magenta field lines. Yellow field lines with footpoints in the southern hemisphere that form high-reaching arches that pass both near the top of the helmet streamer and close to the edge of the pseudostreamer. Figure \ref{fig:helmet}(b) shows the twist gained by the high-reaching yellow field lines, which are in the process of lifting off from the top of the helmet streamer (see also the accompanying animation). The figure also shows that the CME has three footpoint regions. The first, labeled FP1, is near the positive footpoint of the erupting flux rope. The second, labeled FP2, is south of the pseudostreamer and forms when field lines within the CME connecting to FP1 are drawn into the flare current sheet. The resulting interchange-like flare reconnection shifts these CME footpoints from FP1 to FP2. The final footpoint region is FP3, consisting of the remote footpoints of the flux blown out from the helmet streamer. 

Figure \ref{fig:helmet}(c) shows the completely blown-out top of the helmet streamer several hours later. Several hours later still, Figure \ref{fig:helmet}(d) shows the reformation of the helmet streamer. The helmet streamer reforms as a new section of heliospheric current sheet forms following the upward stretching of the blown-out helmet streamer flux. This is exactly analogous to the process by which slow streamer-puff, or stealth, CMEs are thought to form except that in this case the magnetic stress was injected into the helmet streamer flux completely in the corona rather than at the photosphere \citep[e.g.][]{Lynch2016}. 

\begin{figure*}
\centering
\includegraphics[width=\textwidth]{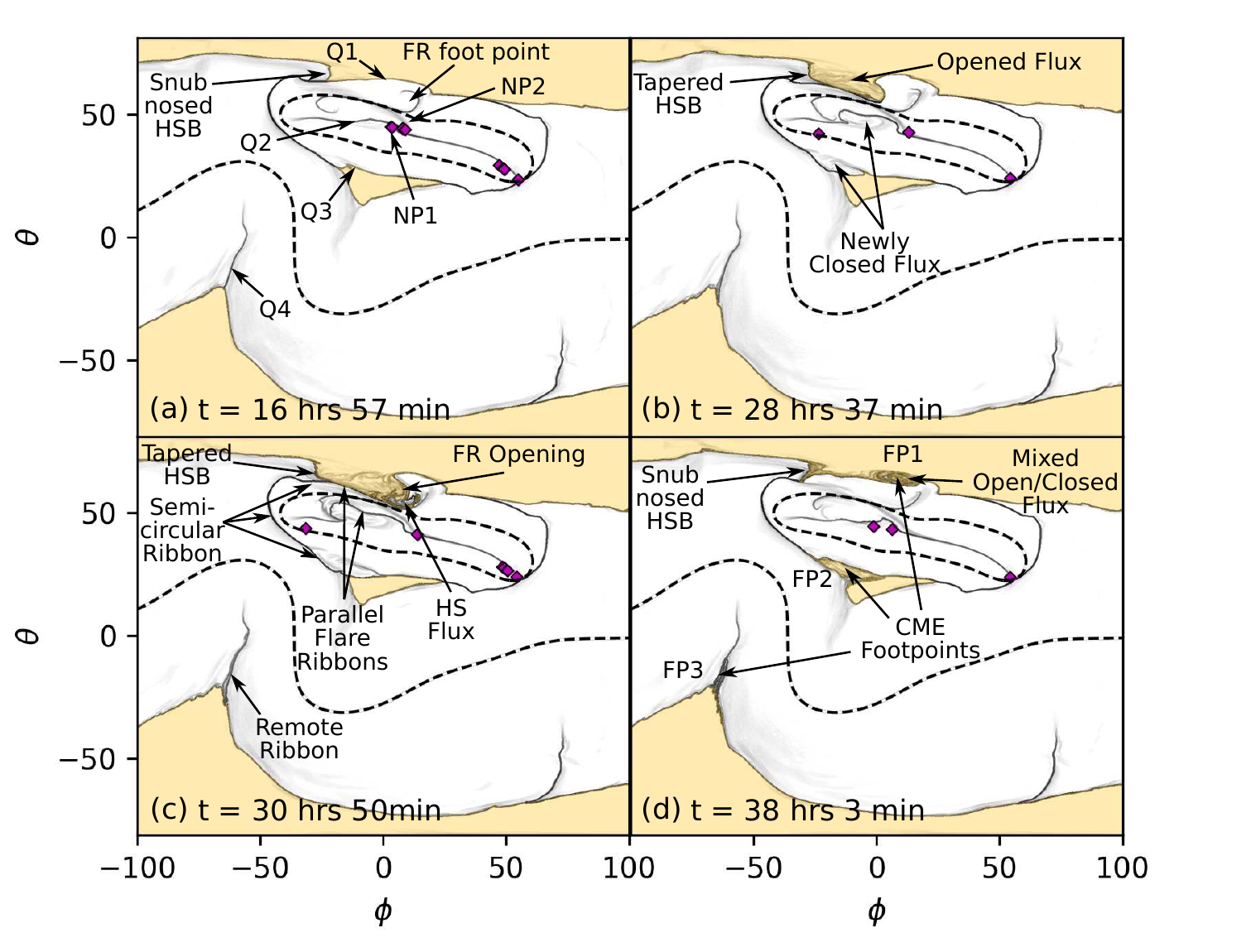}
\caption{Connectivity evolution of the coronal magnetic field. Grey shading shows $\log(Q)$; open and closed fields are shaded yellow and white, respectively; the dashed line shows the PIL; and projected positions of the null points are shown with purple diamonds. HSB = helmet streamer boundary, HS flux = helmet streamer flux, FR = flux rope, FP = CME footpoints. An animation of this figure is available {{showing the evolution. The animation duration is 12 s.}}}
\label{fig:qmap}
\end{figure*}

\begin{figure*}
\centering
\includegraphics[width=\textwidth]{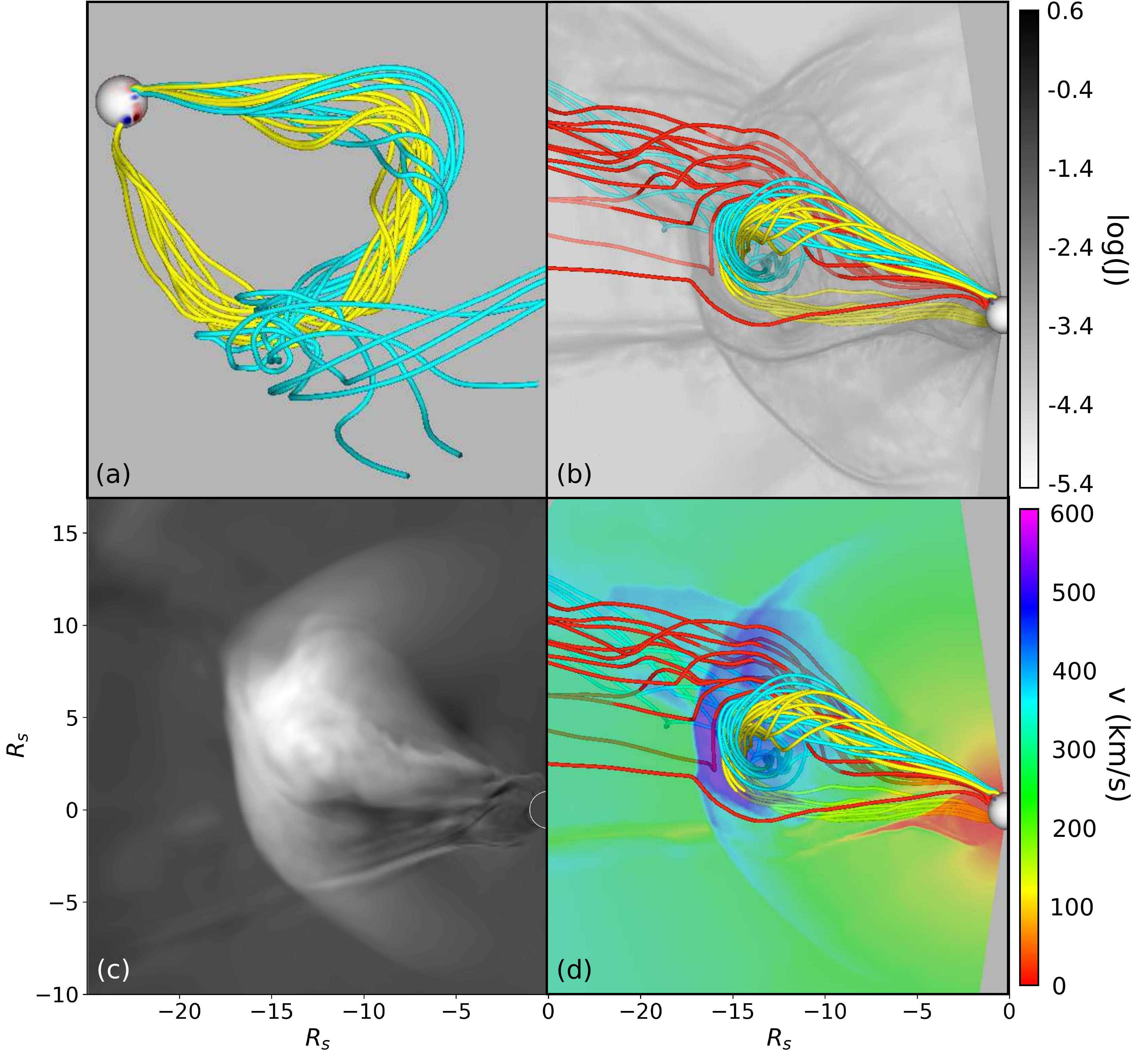}
\caption{(a) Selected CME field lines at $t = 36$\,hrs $23$\,min. (b) \& (d) {{Log of current density (statamp cm$^{-2}$)}} and plasma velocity in a plane across the CME flux rope. (c) Base difference (from $t=0$) of scattered white light using the method employed by \citet{Lynch2016}. An animation of {{panel (d) is available online showing the evolution. The animation duration is 6 s.}}}
\label{fig:cme}
\end{figure*}

\begin{figure*}
\centering
\includegraphics[width=\textwidth]{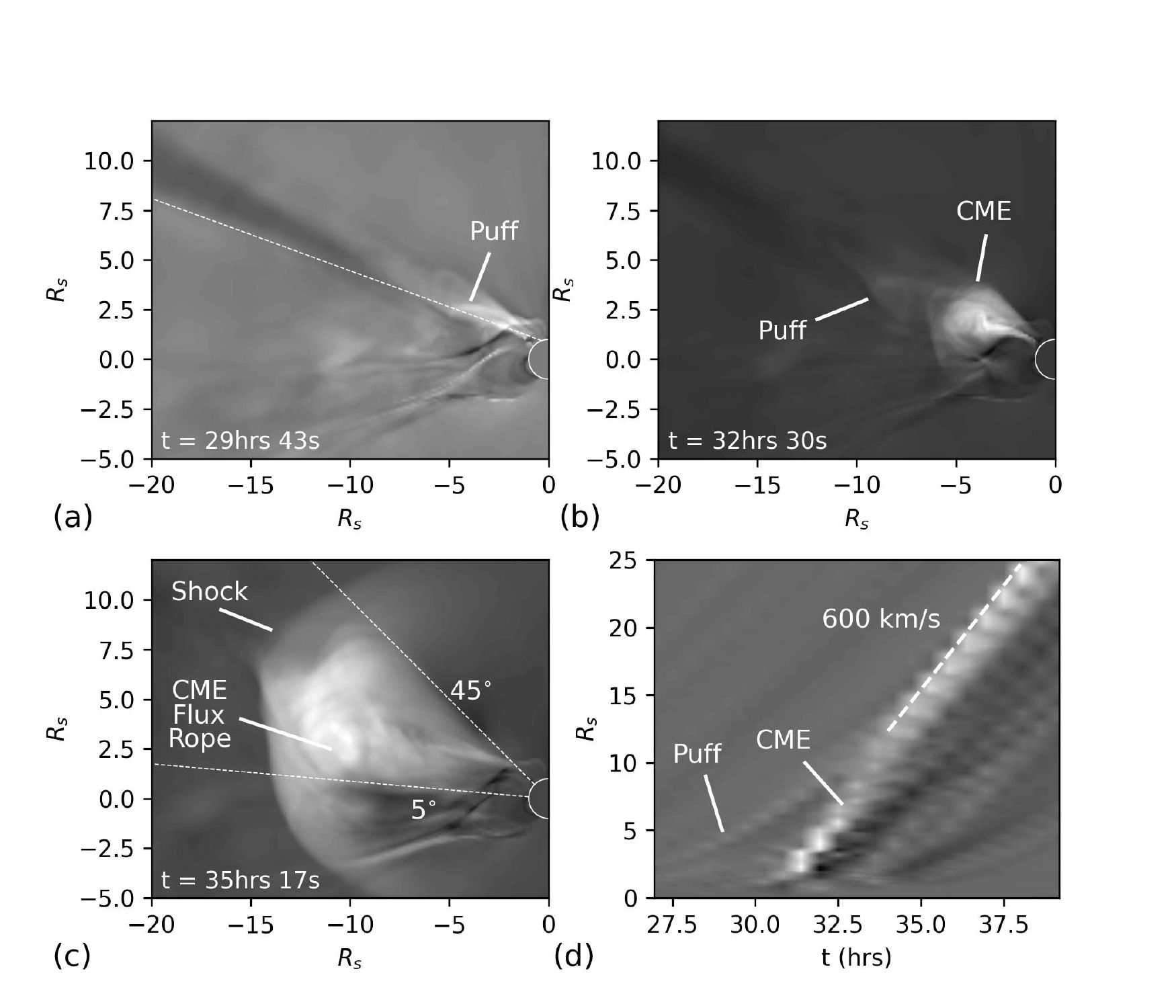}
\caption{(a)-(c) Base-difference images (from $t=0$) of scattered white light. (d) Height-time plot using running difference images. The dashed line in (a) shows the slit for the height-time plot. An animation of this figure is available online {{showing the evolution. The animation duration is 4 s.}}}
\label{fig:cme_obs}
\end{figure*}

\section{Surface Connectivity and Null Evolution}
\label{sec:qmap}
A great deal of insight into this event can be gained by studying the surface connectivity of the magnetic field, which essentially maps the complex 3D evolution of the eruption onto a 2D plane. All the key changes of the global magnetic evolution are captured thereby. At the same time, the connectivity maps are not always intuitive to work with and require some interpretation, which we now provide. 

In Figure \ref{fig:qmap}(a) we show the connectivity prior to breakout initiation. As before, the open-field coronal holes are shown in yellow, magnetically closed regions in white, and $\log(Q)$ in grey scale. In addition to the connectivity, we also located the null points in the low corona (below $3$\,R$_{s}$) using the tri-linear method \citep{Haynes2007,Wyper2016}. The positions of the nulls identified at each time are projected radially onto the surface and shown with diamonds.

As mentioned previously in \S \ref{sec:driving}, during the early driving phase NP1 migrates towards the center of the pseudostreamer (see the accompanying animation), evidently due to the changing shape of the pseudostreamer separatrix as it expands. NP1 resides near NP2 and is in the open field prior to the breakout phase, Figure \ref{fig:qmap}(a). This is confirmed by the snub-nosed shape of the nearby helmet-streamer boundary, indicating that a very narrow corridor of flux is present between the two coronal holes. In fact, NP1 and NP2 (or the localised clusters of nulls that sometimes form from them) actually annihilate each other both before and after the eruption, then finally return via bifurcation. Nulls NP3, NP4, and NP5 are far removed from the filament channel (on the right) and are not involved in the eruption, although they also interact with each other over time. Four curves of high $Q$ are highlighted (Q1 to Q4) that are associated with the fan separatrix footprint of NP1 (Q1 and Q3) and the inner and outer fan of NP2 (Q2 and Q4, respectively). 

Figure \ref{fig:qmap}(b) shows the connectivity well into the breakout phase. In 2D, the progression of breakout reconnection is followed by the motion of the footpoints of separatrices on either side of the filament channel towards each other as strapping field is removed. Meanwhile, the footpoints outlining the lobe regions move apart as strapping field is added \citep[e.g.][]{Lynch2008}. The same progression is evident here in 3D: Q1 and Q2 move towards one another as Q2 and Q3 move apart. At this time, NP1 has migrated back towards the left flank of the pseudostreamer. This shift of the null, which is a natural consequence of the round trip it makes as it moves from the breakout to flare current sheets and back to the center of the pseudostreamer, is the crucial ingredient that leads to the blowout of the helmet streamer. As NP1 moves to the flank, the helmet streamer becomes draped over the pseudostreamer once more and connects to the middle null (NP2). This manifests in the connectivity plot as a return to a tapered, rather than snub-nosed, end to the northern helmet-streamer boundary. The twist liberated by the reconnection of the flux rope is now able to access closed field beneath the helmet streamer. 

Figure \ref{fig:qmap}(c) shows the connectivity as the flux rope reconnects. The northern footpoint of the flux rope has now moved outside the pseudostreamer separatrix and connects to a mixture of open coronal-hole field and closed helmet-streamer flux that was draped over the pseudostreamer. {{At the moment of flux-rope opening, a burst of explosive reconnection is initiated at NP1. Bright local and remote ribbons would be expected to form near the inner and outer spines at Q2 and Q4, as well as along the semi-circular ribbon where the fan plane meets the surface, highlighted in Figure \ref{fig:qmap}(c). The ribbon is semi-circular rather than a closed loop in this case due to the multiple nulls in the pseudostreamer topology. Similar ribbon structures have been observed by, e.g., \citet[][]{Kumar2020}.}} 
Immediately after the flux rope opens and fast flare reconnection is underway, Q1 and Q2 then outline the positions of the main parallel flare ribbons. Subsequently, Q1 and Q2 move apart as NP1 moves back towards the center of the pseudostreamer and into the open-field region, Figure \ref{fig:qmap}(c) and (d). 

The above description reveals that the two key ingredients that lead to the sympathetic blowout of the helmet streamer are (1) the proximity of the edge of the helmet streamer to the edge of the pseudostreamer and (2) the round-trip evolution of the null point as the flux rope reconnects. The shift of NP1 into and back out of the closed field under the helmet streamer seems to be a consequence of both reconnection at the null, as described by \citet{Edmondson2009}, and a shift of the helmet streamer itself driven by interchange reconnection along its flank, similar to that shown by e.g. \citet[][]{Higginson2017}.

Finally, the connectivity map also highlights the three CME footpoints, Figure \ref{fig:qmap}(d). This makes it immediately clear that the southern footpoint FP3 forms next to the original footpoint of the outer spine of NP1. As was highlighted by the field-line evolution in Figs. \ref{fig:breakout} and \ref{fig:helmet}, the northern footpoint of the flux rope disconnects from the CME as it is swept up by the flare reconnection, to reside back in the closed field beneath the pseudostreamer. As can be seen in the animation accompanying Figure \ref{fig:qmap}, FP2 forms when the northern flare ribbon (Q1) sweeps over the positive footpoint of the pre-eruption flux rope. This is the surface signature of the interchange-like flare reconnection that occurs above, shifting some of the CME field-line footpoints from FP1 to FP2. The remaining sheared strapping field adjacent to the pre-eruption flux rope also opens but is not swept over by Q1, thereby forming FP1 at later times (see also Fig.\ \ref{fig:helmet}, cyan field lines). 

\begin{figure*}
\centering
\includegraphics[width=\textwidth]{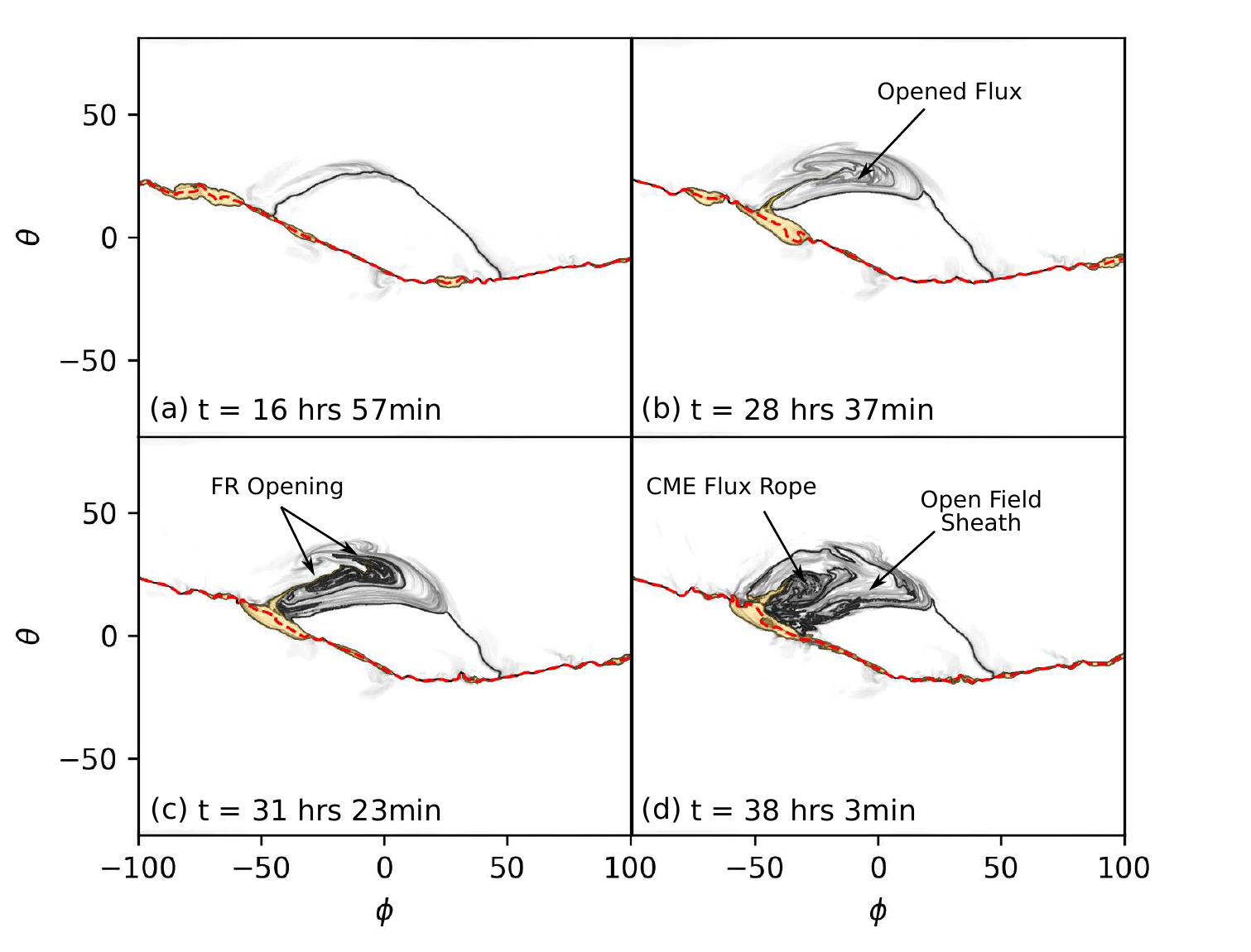}
\caption{Connectivity evolution of the arc {{at $30R_s$}}. Grey scale: $\log(Q)$. White: open coronal-hole field. Yellow: disconnected field lines. The dashed red line shows the reversal in $B_r$ within the heliospheric current sheet. FR = flux rope. An animation of this figure is available online {{showing the evolution. The animation duration is 12 s.}}  }
\label{fig:qmap_top}
\end{figure*}

\section{Coronal Mass Ejection}
\label{sec:cme}
The CME that results is a mixture of twisted/kinked open field lines from the coronal holes and closed field lines blown out from the helmet streamer. They are combined within two main propagating magnetic structures that form the CME. The first is the blown-out helmet-streamer flux, which has footpoints predominantly at FP1 and FP3. Figure \ref{fig:cme}(a) shows that field lines traced from FP3 (yellow) and FP1 (cyan) wrap around one another producing what normally would be identified as the CME flux rope; we will designate it as such from now on. However, the analysis of the previous sections has shown that this flux rope is {\emph{not}} the same flux rope that originally erupted, but rather forms later in the evolution. The original flux rope due to the filament channel was ejected as the pseudostreamer jet; this later flux rope forms from the coupling with the helmet streamer field and erupts as a CME.  When viewed from the side, this CME flux rope appears circular despite being composed of a mixture of open and closed field lines, Figure \ref{fig:cme}(b). This circular feature is embedded within kinked/twisted field lines that form the rest of the CME and have footpoints predominantly at FP2, shown with red field lines in Figure \ref{fig:cme}(b). To more directly compare the structure to observations, a synthetic white-light base-difference image produced at the same viewing angle as the simulation cut is shown in Figure \ref{fig:cme}(c). A bubble-like circular feature is indeed embedded within the broader body of the CME; see also the movie accompanying Figure \ref{fig:cme_obs}. 

\citet{Kumar2020} reported a series of observations showing that, ahead of the CMEs produced by pseudostreamer eruptions, a small jet-like puff first appears, associated with the opening of the strapping field by breakout reconnection shortly before the eruption. In Figure \ref{fig:cme_obs}(a)-(c) we show a series of synthetic coronagraph images from the same viewpoint, just prior to the CME and as it propagates outward. As in the observations, a puff formed by the breakout reconnection outflow and the opening of the strapping flux precedes the main CME. 

Aside from the shock, the CME has a relatively narrow angular extent of about $40^{\circ}$. Additionally, the CME is deflected towards the heliospheric current sheet as it propagates outward, as shown by the curved shape of the CME in Figure \ref{fig:cme_obs}(b). Such a deflection is to be expected, given that the jet straddles the open-closed helmet-streamer boundary and, therefore, is guided towards its apex. We note that this deflection is not the result of magnetic pressure gradients acting on the erupting structure in the low corona, as has been proposed to account for CME deflections in some observed events \citep[e.g.][]{Shen2011b,Kay2013}. In that scenario the erupting magnetic structure is presumed to be a flux rope connected at both ends to the photosphere with an axis at its apex that is roughly perpendicular to the ambient field it encounters. Here the erupting structure propagates \emph{along} field lines in the low corona. The circular feature in the synthetic white-light images appears only after the disturbance reaches the helmet streamer apex; when the CME flux rope forms and the helmet streamer blows out. Prior to this the evolution is more jet-like. Based on the location of the circular feature in Figure \ref{fig:cme_obs}(c) compared with the initial radial trajectory of the jet, we estimate an apparent deflection of around $30^{\circ}$. 

Figure \ref{fig:cme_obs}(d) shows a time/distance plot (J-map) created from running different images evaluated along the line shown in Figure \ref{fig:cme_obs}(a). We chose this non-radial line to capture both the puff and nose of the CME once it is deflected. The pre-eruption puff is simply advected by the ambient wind and follows a slowly accelerating profile. The CME, once launched, quickly overtakes the puff before stabilizing at a roughly constant speed of about $600$\,km s$^{-1}$. Similar CME speeds and acceleration profiles have been measured for CMEs from pseudostreamers \citep[e.g.][]{Wang2018}.

\section{S-Web Connectivity \& Impulsive SEPs}
\label{sec:sep}

The mixture of open and closed flux within the CME provides a natural avenue of escape for impulsive SEPs produced by the flare. Although a full investigation of the expected SEP signatures of this simulation is well beyond the scope of this investigation, we can still make some preliminary remarks based on the connectivity of the CME. There are two potential scenarios where we might expect SEPs to be impulsively accelerated and released in this eruption. The first is the scenario explored by \citet{Masson2013,Masson2019}, whereby flare reconnection {\it{prior}} to reconnection of the filament channel flux rope stores high energy particles within the flux rope. These are then promptly released when the flux rope is opened by reconnection with external field, in this case when the flux rope reaches the breakout current layer. The second is that once the flux rope opens the flare reconnection transitions to interchange-like reconnection at NP1, directly accessing open field lines along which SEPs could escape in a manner similar to a jet. These two scenarios overlap at the moment the flux rope opens, but the latter would continue beyond the initial flux rope reconnection making them potentially distinguishable observationally.

Figure \ref{fig:qmap_top} shows $\log(Q)$ at $30R_{s}$ at various stages throughout the eruption. The times are matched to those in Figure \ref{fig:qmap} for comparison. Regions of disconnected flux within the heliospheric current layer are shown in yellow. Broadly speaking, the open field affected by the eruption is localised around the left half of the S-Web arc, the section which resided above the filament channel. This is a useful result in itself and provides a predictive tool for where impulsively accelerated particles and subsequently the CME itself could be measured in-situ for CMEs originating from pseudostreamers. 

More specifically, throughout the breakout phase the affected section of S-Web arc moves steadily southward as strapping field is opened, Figure \ref{fig:qmap_top}(b). The opening of the flux rope is shown in Figure \ref{fig:qmap_top}(c) by the complex region of high $Q$, which stretches almost to the middle of the arc. Both SEP scenarios would be expected to launch SEPs into this region. This shows that impulsive SEPs could potentially reach far from the heliospheric current sheet, in this case reaching around $30^{\circ}$ from it in latitude and covering a range of around $70^{\circ}$ in longitude, due to the rapidly varying connectivity along the arc. Future work using for example test particles would be required to test this claim definitively. 

Finally, for completeness Figure \ref{fig:qmap_top}(d) shows the connectivity once the helmet streamer has blown out and the CME is fully developed. The circular region corresponds to the open field lines that wrap into the CME flux rope (cyan field lines, Fig.\ \ref{fig:cme}). This is bordered on one side by disconnected field (yellow) within the heliospheric current sheet, and a semi-circular sheath of open field corresponding to the rest of the kinked field lines within the CME (red field lines, Fig.\ \ref{fig:cme}).

\section{Discussion and Conclusions}
\label{sec:disc}

In addition to observations of SEPs, our results presented here on coupled pseudostreamer/helmet-streamer eruptions have clear implications for understanding solar eruptions. First, we emphasize that the assumed magnetic configuration of a pseudostreamer close to the helmet streamer, Figure \ref{fig:grid}, is quite common on the Sun. The coronal hole pattern of Figure \ref{fig:relax} in which a small coronal hole extension is separated from the main coronal hole by a large parasitic polarity region can be seen frequently in source surface maps and is reflected in S-Web maps \citep[e.g.][]{Antiochos2011,Wang2012,Crooker2012,Titov2012}. Furthermore, filament channels are very often observed to form over the PIL of the parasitic region, leading to a jet-like eruption \citep[e.g.][]{Filippov2013,Filippov2015,Yang2015,Wang2018,Kumar2020}, so the scenario described in this paper should be readily observed. We note that our scenario shares some general features with streamer blowouts driven by eruptions beneath the streamer, but near its edge \citep[e.g.][]{Moore2007,Lugaz2011,Panesar2016}. The key difference is that in our scenario the pseudostreamer has a substantial presence outside the streamer, within the coronal hole. 

Another key point is that our jet eruption is somewhat special from a theoretical viewpoint in that it occurs in a 3D system with multiple nulls and separator lines. There have been many simulations of jets and CMEs in the ubiquitous topology of a single null, but to our knowledge, this is the first simulation of a filament channel driven eruption in a multi-null/separator topology. This topology allows for more copious reconnection, because the breakout current sheet  now corresponds to a deformed line segment rather than a deformed point. As argued in \citet{Antiochos1999}, the breakout mechanism requires reconnection to operate, but the reconnection must not be too ``easy'' if the system is to be explosive. Our simulation verifies that breakout can produce explosive eruption even in a general pseudostreamer topology.

The eruption from the actual pseudostreamer, however, is clearly a jet rather than a CME. We do not see closed pseudostreamer flux expanding outward into the heliosphere. This results holds even though the pseudostreamer of Figure \ref{fig:relax} is quite large and, as discussed above, the breakout reconnection is extended, more easily allowing  a flux rope to escape. Instead, the pseudostreamer flux rope is completely destroyed by interchange reconnection either with the open field or the helmet streamer field. We conjecture that this, in fact, is the general result for an eruption from a unipolar background as in a pseudostreamer or jet. The reasons for this are that first, a plasmoid in a unipolar background must have its twist component anti-parallel to the background field on one side or the other, so that reconnection is inevitable. Second, the amount of flux in the pseudostreamer rope is limited by the flux of the parasitic polarity, which is generally small compared to the background. However, if the flux rope can survive out to the Alfv\'en radius $\sim 10R_{s}$, then any interchange beyond this point has no effect on the amount of escaping flux, because a closed field line will still not retract back down to the corona. Depending on the balance between gas dynamic pressure and magnetic tension, this result may hold even at somewhat lower radii, say $5 R_{s}$ or so. These arguments imply that in order for a pseudostreamer eruption to produce a CME when the pseudostreamer is deep within a coronal hole, the parasitic polarity must have an exceptionally large amount of flux and the eruption must develop a high speed early on, which requires that the filament channel be highly sheared. Further simulations are required in order to quantify the fluxes and speeds required for an isolated pseudostreamer CME, if even possible, and observations of pseudostreamer eruptions far from helmet streamers are needed to test the conclusions. 

The situation is very different for a helmet-streamer eruption. In this case the background field is bidirectional and parallel to the plasmoid on both sides. As a result, interchange reconnection does not occur and a plasmoid is free to propagate outward so that even small plasmoids can survive indefinitely in the heliospheric current sheet \citep{Higginson2018}. {{Evidence for this behavior can be gleaned from our simulation in Figures \ref{fig:helmet} and \ref{fig:cme} and their associated animations, in which twisted, closed helmet-streamer field lines are swept up in the CME and expand outwards within or adjacent to the field reversal that defines the heliospheric current sheet. A key conclusion from these figures and animations}} 
is that not only does the field of the CME consist of a mixture of closed and open flux, but the plasma as well consists of a mixture of material that was in the large outer loops of the helmet streamer and the small inner loops of the parasitic polarity. This result has important consequences for in-situ measurements of plasma composition. Even in the outer layers of a streamer-blowout event, the plasma may have originated from the small closed loops of a decayed active region, with a different FIP bias and freeze-in temperature from the helmet streamer. Again, this result requires definitive testing against data. 

This work also has important implications for understanding S-web dynamics. The bulk of the S-Web consists of high-$Q$ arcs that start and end on the heliospheric current sheet, as in Figure \ref{fig:qmap_top}. All these arcs are due to large parasitic polarity regions that produce narrow or singular connections between coronal holes \citep{Antiochos2011,Titov2012,Scott2018,Scott2019}. These polarity regions have PILs, which will invariably become sheared with time; consequently, the type of eruption that we calculated above is bound to be a frequent driver of the S-Web. Our results show that the S-web is intrinsically coupled, in that strong dynamics at one location are likely to lead to dynamics elsewhere with the subsequent mixing of plasma and field lines. This conclusion may help explain the long-standing puzzle of flare SEPs with large longitudinal extent \citep[e.g.][]{Dresling2012,Richardson2014} although further work is required to confirm this.

Perhaps the most important and far-reaching conclusions from this new type of coupled eruption are for understanding the interplay between solar eruptions and the global coronal magnetic field. Taken alongside previous work, our results suggest that CMEs originating nearby the helmet streamer separatrix, be it beneath the helmet streamer or adjacent to it in a coronal hole, should be expected to be deflected, have a more complex morphology and to involve a mixture of open and closed magnetic field lines. Whereas eruptions from deep within the closed field will form classic CMEs and if our conjecture is correct those from pseudostreamers deep within coronal holes will form jets. Accurate global field modeling, supported by further simulation studies of the coupling between eruptions and the global coronal field, is clearly then crucial to predicting and interpreting the in-situ and remote-sensing observations being made by current and future solar missions.

\acknowledgments
It is a pleasure to thank Sophie Masson and Aleida Higginson for useful discussions regarding pseudostreamers and the solar wind. PFW was supported in this work by the award of an RAS fellowship and PROBA2 Guest Investigator grant. PFW thanks Mat West for supplying the SWAP image and Marilena Mierla for support during his visit to ROB. BJL, PK, JTK, CRD and SKA were supported by NASA's H-ISFM, H-LWS, and H-SR programs. Computer resources for the numerical calculations were provided to CRD by NASA's High-End Computing program at the NASA Center for Climate Simulation.

%\bibliographystyle{apj}
%\bibliography{biblionew}
%\end{document}

\end{document}